\documentclass[journal]{new-aiaa}
\makeatletter\if@twocolumn\PassOptionsToPackage{switch}{lineno}\else\fi\makeatother

\usepackage{graphicx,amsmath,amsfonts,amssymb,amstext,amsbsy}
\usepackage{longtable,tabularx,tabulary,xcolor,soul}
\usepackage{rotating}
\usepackage{tikz}
\usepackage{lineno}
\usepackage[T1]{fontenc}
\usepackage[utf8]{inputenc}

\usepackage{url,multirow,morefloats,floatflt,cancel,tfrupee}
\makeatletter

\AtBeginDocument{\@ifpackageloaded{textcomp}{}{\usepackage{textcomp}}}
\makeatother
\usepackage{colortbl}
\usepackage{xcolor}
\usepackage{pifont}
\usepackage[nointegrals]{wasysym}
\makeatletter

\def\mcWidth#1{\csname TY@F#1\endcsname+\tabcolsep}

\def\cAlignHack{\rightskip\@flushglue\leftskip\@flushglue\parindent\z@\parfillskip\z@skip}
\def\rAlignHack{\rightskip\z@skip\leftskip\@flushglue \parindent\z@\parfillskip\z@skip}

\@ifundefined{etal}{}{}

\usepackage{ifxetex}
\ifxetex\else\if@twocolumn\@ifpackageloaded{stfloats}{}{\usepackage{dblfloatfix}}\fi\fi

\AtBeginDocument{
	\expandafter\ifx\csname eqalign\endcsname\relax
	\def\eqalign#1{\null\vcenter{\def\\{\cr}\openup\jot\m@th
			\ialign{\strut$\displaystyle{##}$\hfil&$\displaystyle{{}##}$\hfil
				\crcr#1\crcr}}\,}
	\fi
}

\AtBeginDocument{%
	\@ifpackageloaded{endfloat}%
	{\renewcommand\efloat@iwrite[1]{\immediate\expandafter\protected@write\csname efloat@post#1\endcsname{}}}{\newif\ifefloat@tables}%
}%

\def\BreakURLText#1{\@tfor\brk@tempa:=#1\do{\brk@tempa\hskip0pt}}
\let\lt=<
\let\gt=>
\def\processVert{\ifmmode|\else\textbar\fi}

\@ifundefined{subparagraph}{
	\def\subparagraph{\@startsection{paragraph}{5}{2\parindent}{0ex plus 0.1ex minus 0.1ex}%
		{0ex}{\normalfont\small\itshape}}%
}{}

\newcommand\role[1]{\unskip}
\newcommand\aucollab[1]{\unskip}

\@ifundefined{tsGraphicsScaleX}{\gdef\tsGraphicsScaleX{1}}{}
\@ifundefined{tsGraphicsScaleY}{\gdef\tsGraphicsScaleY{.9}}{}
\def\checkGraphicsWidth{\ifdim\Gin@nat@width>\linewidth
	\tsGraphicsScaleX\linewidth\else\Gin@nat@width\fi}

\def\checkGraphicsHeight{\ifdim\Gin@nat@height>.9\textheight
	\tsGraphicsScaleY\textheight\else\Gin@nat@height\fi}

\def\fixFloatSize#1{}
\let\ts@includegraphics\includegraphics

\def\inlinegraphic[#1]#2{{\edef\@tempa{#1}\edef\baseline@shift{\ifx\@tempa\@empty0\else#1\fi}\edef\tempZ{\the\numexpr(\numexpr(\baseline@shift*\f@size/100))}\protect\raisebox{\tempZ pt}{\ts@includegraphics{#2}}}}

\AtBeginDocument{\def\includegraphics{\@ifnextchar[{\ts@includegraphics}{\ts@includegraphics[width=\checkGraphicsWidth,height=\checkGraphicsHeight,keepaspectratio]}}}

\DeclareMathAlphabet{\mathpzc}{OT1}{pzc}{m}{it}

\def\URL#1#2{\@ifundefined{href}{#2}{\href{#1}{#2}}}

\def\UrlOrds{\do\*\do\-\do\~\do\'\do\"\do\-}%
\g@addto@macro{\UrlBreaks}{\UrlOrds}

\edef\fntEncoding{\f@encoding}

\makeatother

\newif\ifmultipleabstract\multipleabstractfalse%
\usepackage{float}

\begin{document}

\title{Parametric experimental studies on the shock related unsteadiness in a hemispherical spiked body at supersonic flow\footnote{Part of the reported work were presented and published in AIAA Scitech 2019 Forum, 7-11 January 2019, San Diego, California, https://doi.org/10.2514/6.2019-2316}}
\author{Devabrata Sahoo\footnote{Doctoral Researcher}\textsuperscript{\P}}
\affil{Faculty of Aerospace Engineering, Technion-Israel Institute of Technology, Haifa, 3200003, Israel}
\author{S. K. Karthick\footnote{Post-Doctoral Fellow, AIAA Member\unskip, Corresponding author email: skkarthick@ymail.com}\footnote[5]{Contributed equally to this work as first authors}}
\affil{Faculty of Aerospace Engineering, Technion-Israel Institute of Technology, Haifa, 3200003, Israel}
\author{Sudip Das\footnote{Professor}}
\affil{Space Engineering \& Rocketry, Birla Institute of Technology, Mesra-835215, Jharkand, India}
\author{Jacob Cohen\textsuperscript{\S}}
\affil{Faculty of Aerospace Engineering, Technion-Israel Institute of Technology, Haifa, 3200003, Israel}

\maketitle

\begin{abstract}
	
Experimental studies are carried out to investigate the effects of the geometrical parameters with a drag reducing spike on a hemispherical forebody in a supersonic freestream of $M_\infty=2.0$ at $0^{\circ}$ angle of attack. The spike length $(l/D=0.5,1.0,1.5,2.0)$, spike stem diameter $(d/D=0.06,0.12,0.18)$, and spike tip shapes are varied and their influence on the time-averaged, and time-resolved flow field are examined. When $l/D$ increases, a significant reduction in drag ($c_d$) is achieved at $l/D=1.5$, whereas the variation in $d/D$ has only a minor effect. The intensity of the shock-related unsteadiness is reduced with an increase in {$d/D$ from $0.06$ to $0.18$}, whereas changes in $l/D$ have a negligible effect. The effects of spike tip geometry are studied by replacing the sharp spike tip with a hemispherical one having three different base shapes (vertical base, circular base, and elliptical base). Hemispherical spike tip with a vertical base is performing better by reducing $c_d$ and flow unsteadiness. The dominant Spatio-temporal mode arising due to the shock-related unsteadiness is represented through modal analysis of time-resolved shadowgraph images and the findings are consistent with the other measurements.

\end{abstract}
    
\section{Nomenclature}

{\renewcommand\arraystretch{1.0}
\noindent\begin{longtable*}{@{}l @{\quad=\quad} l@{}}
$\alpha$ & Angle of attack ($^\circ$) \\
$\alpha[\Theta(x,y)]$ & Amplitude from DMD analysis \\
$C_p$ & Pressure coefficient\\
$ C_d $ & Over all drag coefficient\\
$ C_{d,base} $ & Base drag coefficient\\
$ c_d $ & Forebody drag coefficient\\
$ \Delta t $ & Time step size $(s)$\\
$ DMD $ & Dynamic mode decomposition\\
$ D $ & Base body diameter $(mm)$\\
$ d $ & Spike stem diameter ($mm$)\\
$ \epsilon $ & Semi-cone angle ($^{\circ}$)\\
$ f $ & Frequency ($Hz$)\\
$ \gamma $ & Specific heat ratio\\
$ I $ & Image intesity\\
$ l $ & Spike length ($mm$)\\
$ \kappa $ & Pressure fluctuation intensity\\
$ M_\infty $ & Freestream Mach number\\
$ \nu_\infty $ & Freestream kinematic viscosity ($m^2/s$)\\
$ psd $ & Power Spectral Density\\
$ POD $ & Proper Orthogonal Decomposition\\
$ \Phi_1\left(x,y\right) $ & Dominant POD spatial mode\\
$ P_0 $ & Free stream total pressure ($Pa$)\\
$ P_\infty $ & Free stream static pressure ($Pa$)\\
$ P_{rms} $ & Root-Mean-Square pressure ($Pa$)\\
$ \overline P $ & Mean pressure ($Pa$)\\
$ P' $ & Pressure fluctuation ($Pa$)\\
$ Re_D $ & Reynolds number based on base body diameter\\
$ \rho_\infty $ & Freestream density ($kg/m^3$)\\
$ S $ & Surface distance along the forebody ($mm$)\\
$SWTBL$ & shock-wave turbulent boundary layer interactions\\
$t$ & Time ($s$)\\
$\Theta\left(x,y\right)$ & DMD spatial modes\\
$T_0$ & Freestream total temperature ($K$)\\
$T_\infty$ & Freestream static temperature ($K$)\\
$U_\infty$ & Freestream velocity ($m/s$)\\
$\zeta$ & Pressure loading
\end{longtable*}}
  
\section{Introduction}
\begin{figure}[]
	\centering 
	\includegraphics[width=\textwidth]{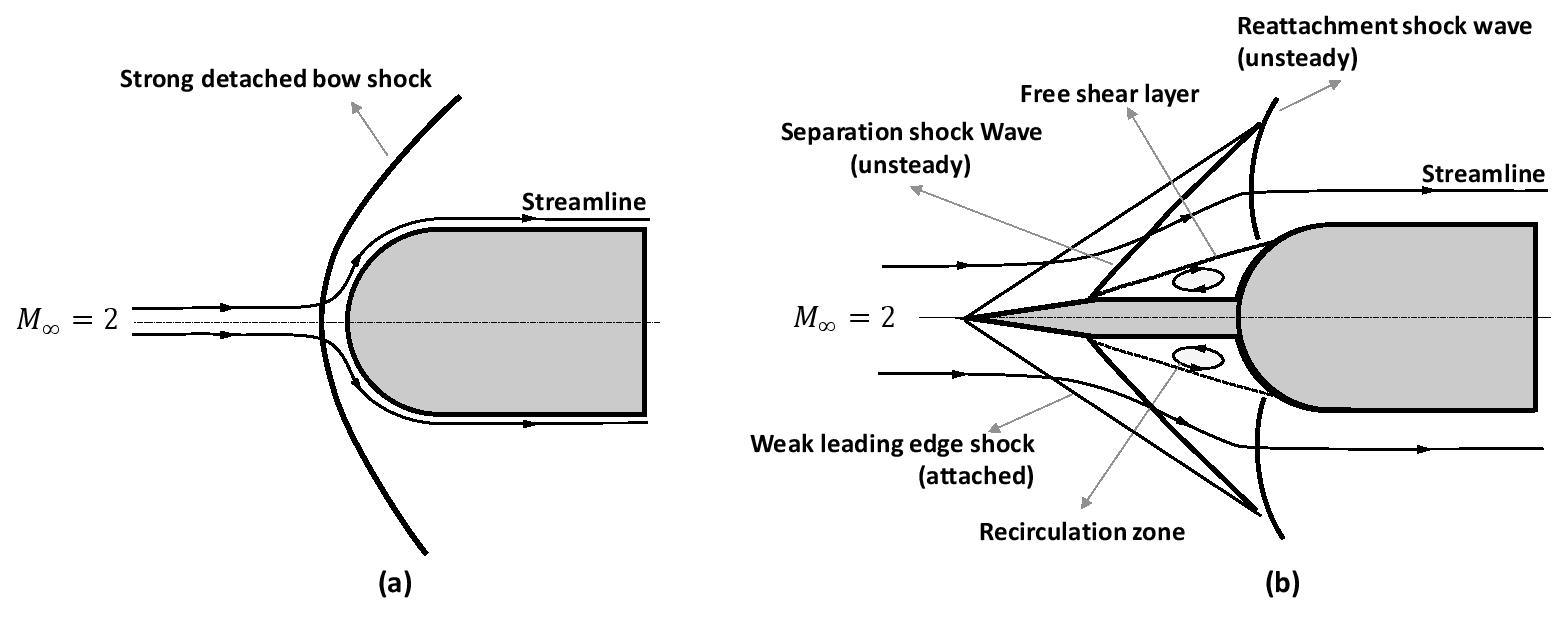}
	\caption{Typical schematic showing the basic flow features observed around a hemispherical forebody (a) without a spike, and (b) with a sharp tip spike, at $M_\infty=2$ and $\alpha=0^\circ$.}
	\label{figure1}
\end{figure}

Hemispherical blunt bodies flying at supersonic speeds find their applications in various categories of aerospace vehicles and missiles \cite{279025:6283319,279025:6283318,279025:10586805,279025:6283326}. The need for housing a payload makes the use of a blunt forebody shape inevitable due to its larger volumetric capability. However, blunt forebodies are exposed to a higher drag and experience aero-heating problems due to the formation of a detached shock wave ahead of them. Among the various means to reduce the encountered wave drag, mounting a slender rod, termed as `spike/aerospike' \unskip~\cite{279025:6283319} at the stagnation point of the blunt-body is commonly preferred for its simplicity. The mounted spike in front of the blunt-body changes the entire flow structure by transforming the single strong bow shock into a system of weaker oblique shock waves, as shown in Figure~\ref{figure1}. In the spike-mounted blunt-body, the flow separates from the spike stem and reattaches near the forebody shoulder. Due to the separation and reattachment of the flow, shock systems are formed near the point of separation (separation shock) and the point of reattachment (reattachment shock). Also, a weak leading-edge shock is formed at the spike tip. The separated flow around the spike stem leads to the generation of a re-circulation region and screens a larger portion of the blunt forebody surface from the external flow. Such an event reduces the strength of the bow shock in front of the forebody without a spike. Consequently, the associated surface pressure distribution and drag are significantly reduced \cite{279025:6283316}.

Successful reduction in the forebody drag coefficient ($c_d$) was achieved in the past by utilizing spikes at both supersonic and hypersonic flow regimes. Alexander \unskip~\cite{279025:6283317} was the first one to suggest the use of an aerospike for drag reduction on blunt bodies experimentally between $1<M_\infty<1.37$. Piland and Putland \unskip~\cite{279025:6283319} are considered to be the first to use the term `spike,' though they did not get any appreciable reduction in $c_d$ in the range of $0.7<M_\infty<1.3$. Later, Jones \unskip~\cite{279025:6283318} conducted an experimental investigations on the flow separation from spike mounted blunt forebodies at $M_\infty=2.72$. {Jones} explained the mechanism and governing criteria of flow separation caused by the mounted spike along with the effect of spike length ($l$). {Jones} also observed that the lowest $c_d$ was achieved for the longest spike that maintained the flow separation at the spike tip. 

{Myshenkov \mbox{\cite{279025:6283320}} used a finite-difference approach to solve the Navier-Stokes equations in a two-dimensional (2D) axisymmetric computational domains to investigate the formation and development of the flow separation caused by a pointed spike mounted ahead of a flat-face cylinder in the range of  $0.5<M_\infty<3.0$.} Myshenkov also varied the spike length (up to $l/D$=3.25) and discussed the roles of $M_\infty$, $Re_D$, and $l/D$ on the size of the separation region formed. The size of the separation region shrinks with a reduction in $Re_D$. Similarly, it expands with an increment in $M_\infty$ until $M_\infty$=1.4. In addition, a non-conical separation region was also seen to be formed along the $l/D$, while changing the $l/D$ from 0 to 2. Paskonov and Cheranova \unskip~\cite{279025:6283321} solved the flow field computationally around a cone cylindrical and flat-face cylindrical models equipped with a pointed aerospike at supersonic speeds. They studied the effect of $l/D$ on the flow structure up to $l/D=1.0$, for which they obtained a maximum drag reduction of 28.5\%. Hutt and Howe \unskip~\cite{279025:6283322} found that increasing the $l/D$ of a forward-facing aerodynamic spike mounted on a family of supersonic blunt cone nosed bodies, reduces $c_d$ only up to a critical value of $l/D$. They used different spike cross-sections and reported that the triangular one showed improved benefits in lift performance. Yamauchi, Fujii and Higashino \unskip~\cite{279025:6283323} also studied numerically the flow field around an aerospike mounted blunt body at $M_\infty$=2.01, 4.14, and 6.80 for different $l/D$=0.5, 1.0 and 2.0. They reported that the area of the re-circulation region formed by mounting a spike increases with spike length ($l$) and thereby reduces the forebody surface static pressure distribution and the associated $c_d$. The influence of $M_\infty$ on the size of the separated region was found to be minimal.

Milicev and Pavlovic \unskip~\cite{279025:6283324} showed that the effectiveness of the aerospikes could be increased further by using a flat-faced or hemispherical-faced spikes called as `aerodisks.' Das et al. \unskip~\cite{279025:6283330} observed a drag reduction of up to 68\% at $M_\infty=2.0$ with spikes having sharp and hemispherical blunt tips. Similar studies were conducted at hypersonic speeds ($M_\infty>5$) and a successful reduction of drag were reported \cite{279025:6283325,279025:6283326,279025:6283327,279025:6283328}. Huang et al.,\unskip~\cite{279025:10477596} has reported the effect of aerodisks on drag as well as heat flux reduction on the hemispherical forebody using numerical simulations at {$M_\infty=4.9$}. They reported a maximum drag reduction of 54.92\% and claimed that the reduction of drag is proportional to the length of the spike ($l$) and the diameter of the aerodisk. However, these parameters have a minimal impact on the static temperature distribution along the forebody surface. Recently, Zhong and Yan \unskip~\cite{279025:10477679}, studied the effect of an aerodisk on drag reduction for an elliptical blunt-body moving at a hypersonic speed of $M_\infty=5.0$ using numerical simulations and reported a maximum drag reduction of 41.9\%.

Along with the reduction of $c_d$, the mounted spike generates severe flow unsteadiness that cannot be ignored as it could lead to a loss of maneuverability and structural failure. Mair \unskip~\cite{279025:6283331} was the first to record the unstable flow phenomena on different spiked blunt bodies (Hemispheres, Flat cylinders, and a 2-dimensional equivalent model) at $M_\infty=1.96$. He reported a significant reduction in the level of flow unsteadiness by changing the flat-face forebody to a hemispherical shape. Maull \unskip~\cite{279025:6283334} also reported the disappearance of the unsteadiness in the hemispherical spiked body configuration at $M_\infty=6.8$. Later, in the 1980s, few of the researchers \cite{279025:6283347,279025:6283348,279025:6283349} measured the pressure fluctuations over a near hemispherical base body resembling the Trident missile model at supersonic speed. They measured large pressure fluctuations (30\% of the dynamic freestream pressure) on the sides of the spike base, especially at transonic speeds. Besides, they also reported higher-pressure fluctuations (42\% of the freestream dynamic pressure) as $\alpha>7^{\circ}$. Other investigators \cite{279025:6283325,279025:6283351,279025:6283352}, conducted numerical studies and also observed unsteady flow over spiked blunt hemisphere bodies at hypersonic speeds. Besides, Sahoo et al., \unskip~\cite{279025:6283353} reported the presence of flow unsteadiness over a spiked hemispherical forebody, based on computations and experiments. They measured the pressure fluctuations at several locations along the hemispherical forebody mounted with spikes of different shapes and sizes at $M_\infty=2.0$. Xue at al., \unskip~\cite{279025:10477806} used detached eddy simulations and captured the two distinct instability modes of flow unsteadiness (oscillation and pulsation) arising over an aerodisk mounted blunt body at $M_\infty=2.0$.

Thus, in the last two decades, many investigators have reported the usage of spiked bodies of different parameters to achieve superior heat transfer and drag reduction capabilities, but only a few have focused on the resulting flow unsteadiness. The review paper of Huang and Chen \unskip~\cite{279025:10477680} lists out many such studies, focusing mainly on drag and heat flux reduction from the computational standpoint. Thus, the forementioned events motivate the authors to perform a detailed experimental parametric study on the blunt spiked bodies addressing the drag reduction and the flow unsteadiness, in particular.

In the present research, parametric studies are carried out using time-averaged and time-resolved experiments over a hemispherical forebody mounted with and without sharp tip spikes of different diameters ($d$) and lengths $(l)$ at a freestream Mach number of $M_\infty=2.0$ at $\alpha=0^{\circ}$. Besides, the effect of the spike tip geometry is also presented by replacing the sharp tip spike with a hemispherical one having three different base shapes (vertical, circular, and elliptical). Following the introduction, a detail description of the experimentation procedure is given, including details about the facility, geometrical aspects of the selected configurations, and a brief layout of the measurement and data analysis methodology. The next section provides details about the operating conditions and the uncertainties observed in the experimentation and data analysis. In the results and discussions section, the effect of spike diameter $(d/D)$, spike length $(l/D)$, and spike tip geometry on the time-averaged and time-resolved flow field are presented and supplemented through modal analysis from the time-resolved shadowgraph images. The last section carries the major conclusions of the present investigations.

\section{Experimentation}
The experiments were carried out in the blowdown supersonic wind-tunnel available at the Faculty of Aerospace Engineering, Technion-Israel Institute of Technology. Details of the facility, model geometry, and measurement methodology adopted in the present work are provided in the following sub-sections.

\subsection{Facility}
\begin{figure}[htb!]
	\centering \includegraphics{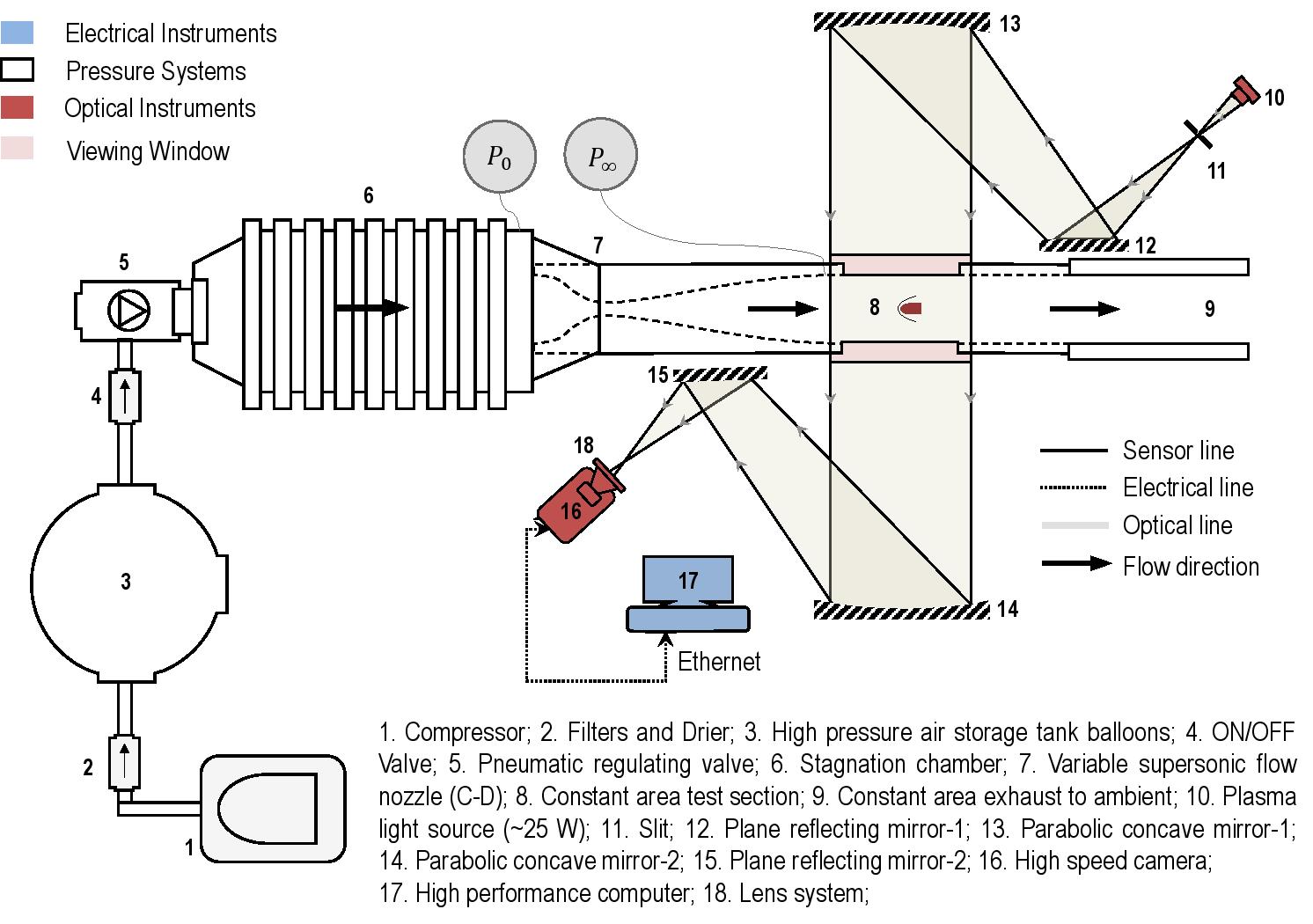} 
	\caption{Schematic of the supersonic wind tunnel with a `Z-type' shadowgraph imaging \protect\cite{settles} setup that is used to study the shock-related unsteadiness in the hemispherical spiked body configurations at $M_\infty=2.0$. Flow is from left to right.}
	\label{figure2}
\end{figure}

Figure~\ref{figure2} shows the layout of the blowdown supersonic wind tunnel facility. The compressed air to the wind-tunnel is drawn from 48 balloons, each having a storage volume of 0.6 $m^3$ at 20 MPa. The storage balloons are charged using a 400 kW, 5-stage piston-type air compressor at a discharge rate of 0.5 $m^3/s$. The tunnel has a fixed test section size of 400 mm (width), 500 mm (height), and 1000 mm (length). It is designed with a variable throat convergent-divergent (CD) nozzle that can generate a freestream flow of Mach number in the range of $1.6<M_\infty<3.5$. Before storing the compressed air, moisture and impurities are removed using appropriate filter systems. A pneumatic control valve controls the compressed air flow entering the stagnation chamber from the air storage. Freestream flow conditions (see Table \ref{table1}) are established in the constant area in the test section. The sidewalls of the test section are mounted with optical quality glass windows (BK-7) (550 mm in length and 300 mm in height) to enable flow visualization studies. Inside the test section, models are mounted on a downstream sting to enable force and pressure measurements. After flowing past the test section, the air is discharged to the ambient surrounding through a constant-area duct having a similar size as that of the test section.

\subsection{Model geometry}
\begin{figure}[htb!]
	\centering \includegraphics[width=0.7\textwidth]{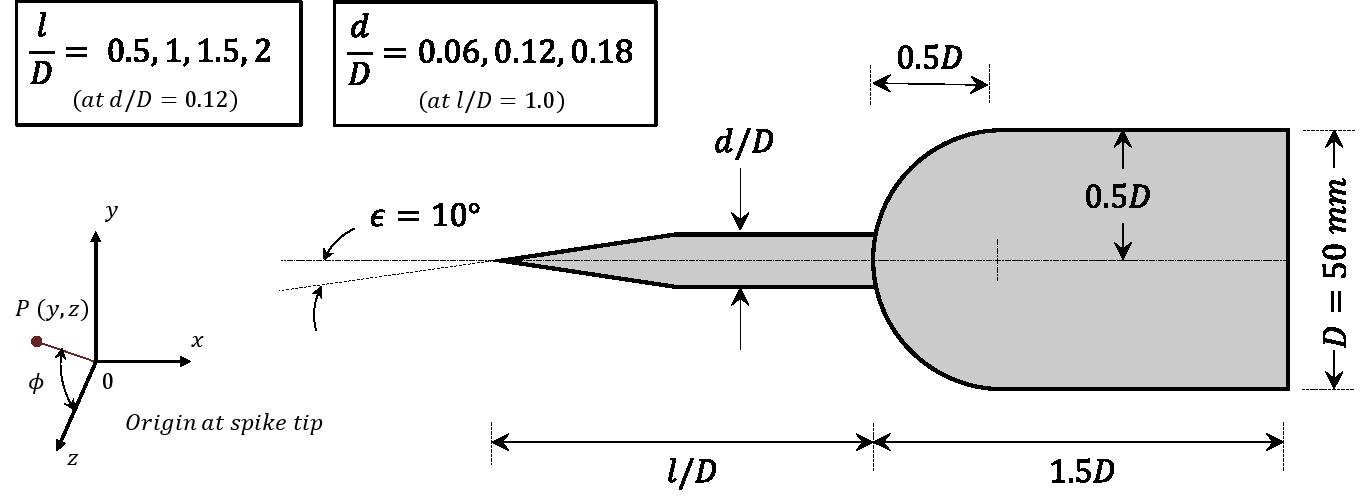}
	\caption{Schematic (not drawn to scale) showing the primary geometrical parameters describing the hemispherical forebody configuration with a sharp tip spike. Different sharp tip spike lengths $(l/D)$ and diameters ($d/D$) adopted in the present experiments are shown. The origin is at the spike tip.}
	\label{figure3}
\end{figure}

A hemispherical blunt-body having a base diameter ($D$) of 50 mm, an overall length of $1.5D$, and a sharp tip spike having a semi-cone angle of $\epsilon=10^{\circ}$, has been utilized for the present study. Spikes with a fixed stem diameter ($d$) of 6 mm $(0.12D)$ are employed to study the effect of spike length $(l/D)$, ranging from [$l/D$]=0.5 to 2.0 (in steps of 0.5). Spikes with a fixed length $(l)$ of 50 mm ($D$) are used to study the effect of spike diameter ($d/D$), ranging from [$d/D$]=0.06 to 0.18 (in steps of 0.06). The primary geometrical details of the configurations utilized in the present study are shown in Figure~\ref{figure3}. The cross-sectional area of the test model results in only 0.98\% blockage in the test section, provided the angle of attack is kept at $0^{\circ}$.

\subsection{Measurement methodology}The flow field has been visualized using the standard `Z-type' shadowgraph technique \unskip~\cite{279025:6389799}. A high-intensity plasma light (white light) source of 25 W has been used to produce the required light intensity for the shadowgraph imaging. The light passes through a slit, which forms the point light source for shadowgraphy. A phantom V211 model, monochrome, 12-bit high-speed camera has been utilized to capture the shadowgraph images.  The time-averaged flow field has been obtained from the high resolution (1280 $\times$ 800) shadowgraph images recorded at a lower frame rate of $f=2200$ Hz and a relatively long exposure time of 5 $\mu s$. A set of 1000 frames has been used to obtain the time-averaged image.

\begin{figure}[htb!]
	\centering \includegraphics[width=0.75\textwidth]{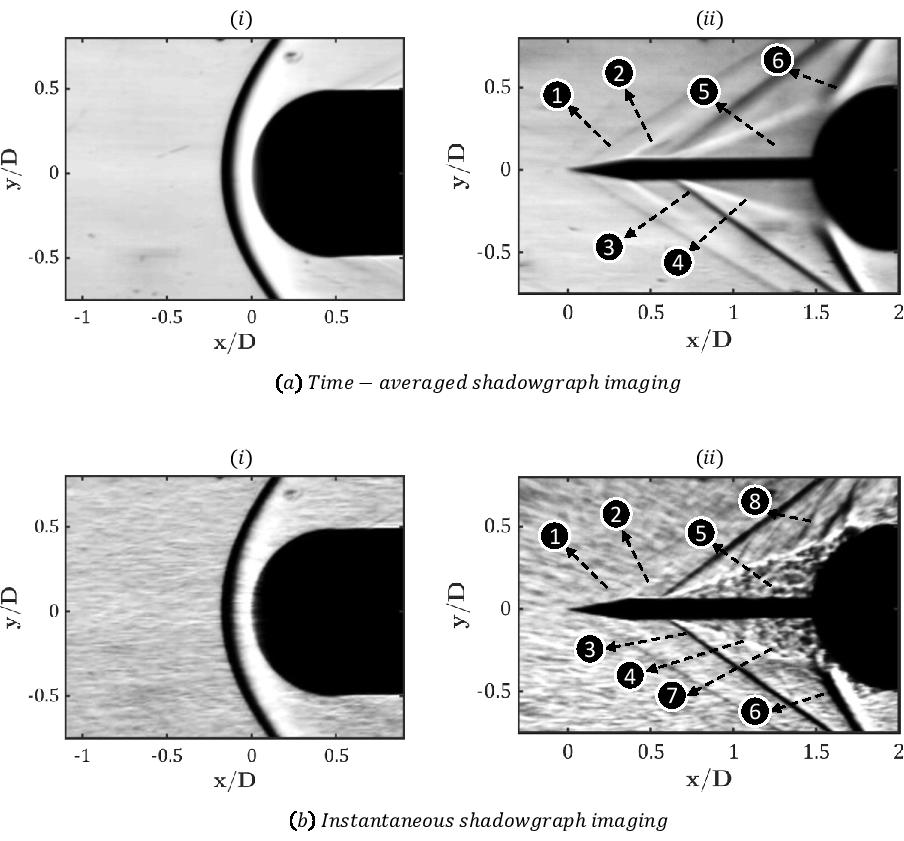}
	\caption{Typical shadowgraph images showing the dominant flow features observed around a hemispherical forebody configuration without (left, i) and with (right, ii) a sharp tip spike $\left(l/D=1.5,\;d/D=0.12\right)$ using (a) time-averaged ($\left\|\bar{\boldsymbol I}\right\|$) and (b) instantaneous ($\left\|{\boldsymbol I}\right\|$) time-resolved shadowgraph imaging at a freestream supersonic flow Mach number $M_\infty=2.0$. Flow features: 1. Weak leading edge shock, 2. Expansion fan, 3. Separation shock, 4. Separated free shear layer, 5. Re-circulation region, 6. Reattachment shock, 7. Large scale structures, 8. Shocklets. In case of the forebody without a spike, a strong bow shock is seen. Flow is from left to right}
	\label{figure4}
\end{figure}

Furthermore, time-resolved flow fields have been obtained at a higher frame rate of $f = 43000$ Hz while keeping the minimal available exposure time of 2 $\mu s$ to capture the shock-related unsteadiness associated with the hemispherical spiked body. Relatively low frame resolution (256 $\times$ 160) has been utilized to record the time-resolved flow field with a higher frame rate and least exposure time. Typical time-averaged and instantaneous\footnote{Corresponding time-resolved shadowgraph video file is given in the supplementary under the name `video7,' and `video5'.} shadowgraph images obtained over a hemispherical forebody configuration without and with a sharp tip spike ($l/D$=1.5; $d/D$=0.12) are shown in Figure~\ref{figure4}a and b, where the dominant flow features are marked. In the case of the forebody without the spike (Figure \ref{figure4}-i), a strong bow shock is seen whose strength could be correlated to the thickness of the shock from the shadowgraph image itself. For a body with a sharp tip spike, as shown in Figure \ref{figure4}-ii, the strong bow shock has disappeared and replaced with a system of weak oblique shocks. The decrement in shock thickness, as observed from the shadowgraph image of Figure \ref{figure4}-ii, is once again a representation of weaker shock strength.

The drag measurements have been carried out on the hemispherical forebody mounted with various spike geometries using an in-house built six-component strain gauge balance. The balance, which has the capability of measuring axial and normal forces of 50 kg and 250 kg, respectively, are powered by a DC voltage of 18 V. The utilized strain-gauge type force-balance system measures the overall drag coefficient ($ C_d $). However, for all practical purposes, the forebody drag coefficient ($ c_d $) is preferred. In the present study, the forebody drag coefficient ($c_d$) is calculated by subtracting the measured base drag coefficient ($ C_{d, base} $) from the measured overall drag coefficient ($ C_d $). For this purpose, a pressure sensor has been mounted at the base of the model to provide the base drag coefficient (\cite{279025:9389372}). All the measured outputs have been acquired using a National Instruments data acquisition system (NI-DAQ), which includes 32 sequential measuring channels. All the analog channels have a sampling rate of 250-kilo samples per second. Before getting transferred to a computer, the acquired signals have been filtered, amplified, and converted into digital datasets.

\begin{figure}[htb!]
	\centering \includegraphics [width=0.8\textwidth] {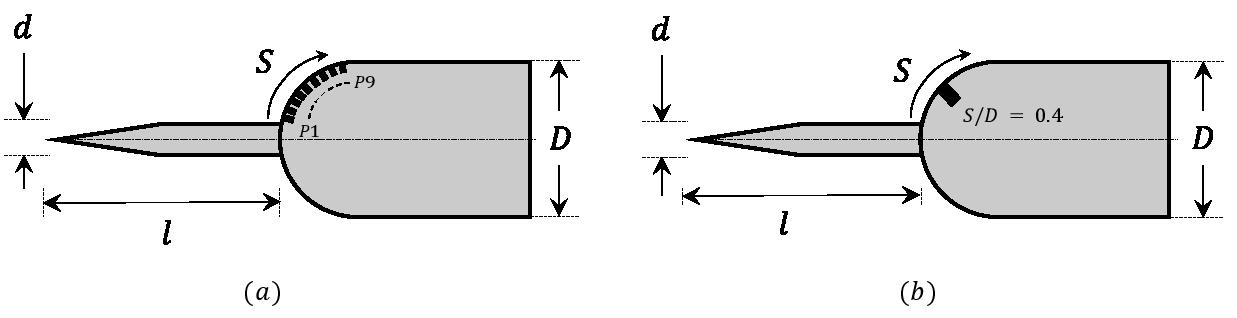}
	\caption{Schematic showing the location of the (a) mean static pressure measurements and (b) unsteady pressure measurement. Here $S$ denotes the distance along the surface of the hemisphere starting from the axis.}
	\label{figure5}
\end{figure}

\begin{figure}[]
	\centering \includegraphics[width=0.5\textwidth]{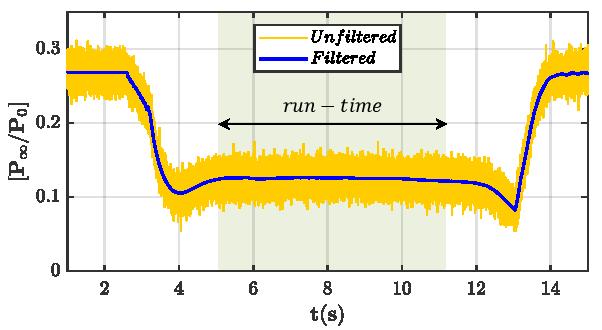}
	\caption{A typical wall static pressure signal (unfiltered and filtered) observed during the supersonic wind tunnel operation ($M_\infty=2.0$, $P_\infty=0.44 \times 10^5\;Pa$) defining the achievable steady test time of about few seconds to acquire all the steady/unsteady signals and images. The physical location of the displayed pressure measurement is shown in Figure \ref{figure2} as $P_\infty$.}
	\label{figure6}
\end{figure}

\begin{figure}[]
	\centering \includegraphics[width=0.5\linewidth]{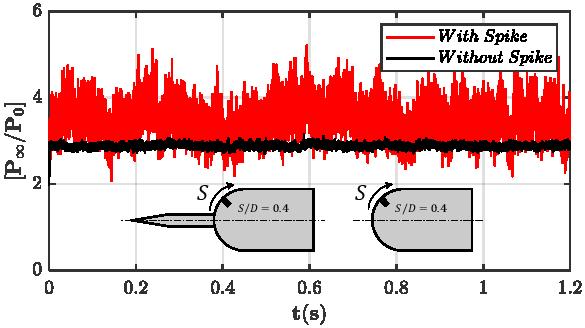} 
	\caption{Typical unsteady pressure fluctuations observed over the hemispherical forebody at $[S/D]$ = 0.4 (line: black-without spike, red-with a sharp tip spike of $[l/D]$=1.5 and $[d/D]$=0.12) at $M_\infty=2.0$.}
	\label{figure7}
\end{figure}

Honeywell static pressure sensors having a maximum range of 50 psi have been used for mean static pressure measurements. The pressure sensors have been powered with a DC voltage of 10 V. The sensors have been mounted at 9 different locations along the hemispherical forebody surface, as shown in Figure~\ref{figure5}a. To measure the unsteady pressure fluctuations generated over the hemispherical forebody surface, an unsteady pressure transducer (XCL-100-50A, Make: Kulite), having a maximum frequency response of 250 kHz has been used. The transducer is flush-mounted near the reattachment point at a surface location of $S/D$=0.4 and $\phi=90^\circ $, as shown in Figure~\ref{figure5}b. Here, `$S$' denotes the distance along the surface of the forebody from the axis (or the stagnation point in the absence of a spike). A sampling rate of 50 kHz for 2 s has been utilized for the present experiments to acquire an unsteady pressure signal. The filtered and unfiltered static pressure data measured at the tunnel wall during the entire run of a typical test can be seen in Figure~\ref{figure6}a. All the experimental data have been acquired during the steady test time, as indicated in Figure~\ref{figure6}a. Typical unsteady pressure signals measured near the shoulder of the hemispherical forebody without and with a sharp tip spike of length, [$l/D$]=1.0, and diameter, [$d/D$]=0.12 are shown in Figure~\ref{figure7}. These pressure signals are then processed to extract the power spectral density ($psd$) for all the cases of hemispherical spiked body configurations used in the present study. 

Modal analysis using POD (Proper Orthogonal Decomposition) and DMD (Dynamic Mode Decomposition) has been carried out on the time-resolved shadowgraph images to supplement the findings from the unsteady pressure measurements. The classic snapshot method of modal analysis has been adopted with the obtained shadowgraph images \cite{279025:6382261,279025:6382262,279025:6382263,279025:6382258,279025:6382259,kutz}. The images have been pre-processed to remove anomalies from the parasite reflections from the glass window or the fluctuations in the utilized light source (\cite{279025:6510007}). Recommendations provided by Rao and Karthick \unskip~\cite{279025:6389893} are followed to minimize the aliasing and exposure effects. Nearly 1000 images have been found to be enough for representing the statistics through modal analysis. The dominant spatial mode ($\Phi_1(x,y)$) are extracted from the POD analysis and the overall dynamic temporal spectra ($\alpha \left[\Theta\left(x,y\right)\right],f$) are extracted from the DMD analysis as detailed in \unskip~\cite{kutz}.

\subsection{Operating conditions}All the experiments have been carried out at a freestream Mach number ($M_\infty$) of 2.0 with a settling chamber pressure ($P_0$) of 3.5 bar. The freestream Reynolds number based on the hemispherical forebody diameter ($Re_D$) is 2.16 $\times 10^6$. Before starting the experiments, necessary wind tunnel calibration for $M_\infty$ has been conducted. From the unsteady pressure measurements at the wall of the test section, the normalized free stream static pressure fluctuations intensity in the supersonic tunnel has been calculated to be 1.57\% (see equation \ref{eq-kappa}) which defines the turbulence level of the wind tunnel. Table~\ref{table1} shows the tunnel operating conditions during the experiments.

\begin{table}[]
	\caption{{Freestream flow conditions achieved in the test section of the blowdown type supersonic wind tunnel during the present investigation.}}
	\label{table1}
	\centering 
	\begin{tabular}{cc}
		\hline 
		\textbf{Quantities} &
		\textbf{Values$^*$}\\
		\hline
		Total Pressure $(P_0,\;Pa)$  &
		$3.48 \times 10{^{5}} {\pm} 5\%$\\
		Total Temperature $(T_0,\;K)$  &
		$294.46 {\pm} 2\% $\\
		Freestream Temperature $(T_\infty,\;K)$  &
		$163.59 {\pm} 2\%$\\
		Freestream Pressure $(P_\infty,\;Pa)$  &
		$0.44 \times 10{^{5}} {\pm} 5\%$\\
		Freestream Density $(\rho_\infty,\;kg/m{^{3}})$  &
		$0.94 {\pm} 5\%$\\
		Freestream Velocity $(U_\infty,\;m/s)$  &
		$515.15 {\pm} 2\%$\\
		Freestream Kinematic Viscosity $(\nu_\infty,\;m{^{2}/s})$ &
		$1.12 \times 10{^{-5}} {\pm} 2\% $\\
		Freestream Mach number $(M_\infty)$ &
		$2.01{\pm}1\%$\\
		Reynolds number based on $D$ ($Re_D=U_\infty D/\nu_\infty$) &
		$2.16 \times 10{^{6}} {\pm} 5\%$\\
		\hline
		\multicolumn{2}{c}{*uncertainty is given in percentage about the measured values}\\ 
	\end{tabular}
\end{table}

\section{Uncertainty}
Data convergence and repeatability have been ensured by repeating each experiment at least 3 times, and the reported data are the ensemble average of the multiple runs. The uncertainty in the unsteady pressure measurement is estimated to be less than $\pm5\% $. Similarly, the uncertainty of the drag coefficient is estimated to be less than $\pm5\% $. The uncertainties in the measurements of other derived quantities in the present work are listed in Table~\ref{table1}. In the modal analysis also, the reported values are the ensemble average of the multiple sets performed for the individual cases. The total uncertainty involved in the spatial mode representation, including the error propagation in the image processing routines and the calibration procedure, is estimated to be around $\pm4\%$. The total uncertainty in the temporal mode representation is estimated to be around $\pm3\%$.

\section{Results and Discussion}

\subsection{Effect of the spike diameter $(d)$ and spike length $(l)$}

\subsubsection{Investigation of time-averaged flow field}
\begin{figure}[htb!]
	\centering \includegraphics[width=0.9\textwidth]{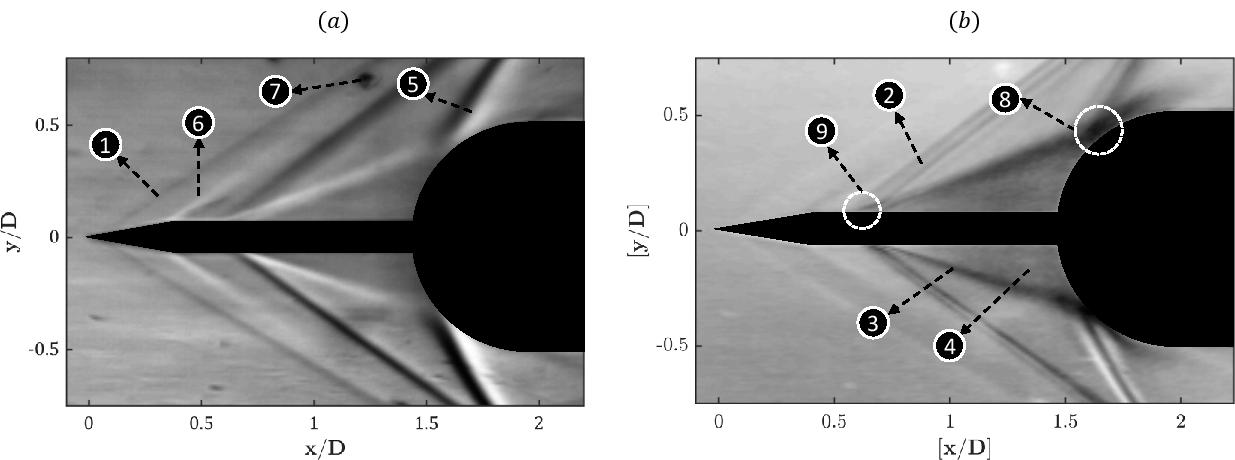} 
	\caption{Typical time-averaged shadowgraph image obtained through the operation of (a)  $\left\|\bar{\boldsymbol I}\right\|$ and (b) $\left\|\bar{\boldsymbol I}-\boldsymbol I_{rms}\right\|$ showing the dominant flow features observed around a hemispherical body with a sharp tip spike $\left(l/D=1.5,\;d/D=0.12\right)$ at $M_\infty=2.0$. Flow features: 1. Weak leading edge shock; 2. Separation shock; 3. Separated shear layer, 4. Re-circulation region; 5. Reattachment shock; 6. Expansion fan, 7. Window defects/image artifacts, 8. Reattachment point, 9. Separation point. Flow is from left to right.}
	\label{figure8}
\end{figure}

Figure~\ref{figure8}a and b show the time-averaged shadowgraph images obtained over a hemisphere with a sharp spike tip of $l/D$=1.5 and $d/D$=0.12 using the operation of $\left\|\bar{\boldsymbol I}\right\|$ and $\left\|\bar{\boldsymbol I}-\boldsymbol I_{rms}\right\|$ at a $M_\infty$=2.0. Features with image artifacts are present in regular operation of $\left\|\bar{\boldsymbol I}\right\|$ as shown in Figure~\ref{figure8}a whereas those have been removed during the operation of $\left\|\bar{\boldsymbol I}-\boldsymbol I_{rms}\right\|$ as shown in Figure~\ref{figure8}b. Besides, the later operation also marks the location of dynamic (unsteady) events. The key flow features are marked in Figure~\ref{figure8}a and b, which includes the weak leading edge shock, the separated shock originated from the spike tip shoulder, the adjacent separated free shear layer bounding the re-circulation region and extending to the hemispherical forebody near the shoulder, where the reattachment shock is formed. Due to the turning of the flow across the separation shock in the presence of a re-circulation region, the flow turn angle required near the forebody is reduced, and consequently, the strength of the shock wave (reattachment shock) at the reattachment region weakens in comparison with the body having no spike (see Figure \ref{figure4}(i)). Later, the flow is turned back to the freestream direction through the formation of a series of expansion fans (not shown in the figure). 

As mentioned earlier, the re-circulation region formed while mounting the spike screens the front surface of the hemisphere from the supersonic freestream flow. A system of weak oblique shock waves forms ahead of the body which results in a change in the pressure distribution over the hemisphere body and the associated wave drag experienced by the body. The change in the flow features, the mean static pressure distribution, and the drag experienced by the forebody while varying the parameters (spike diameter-$d$, and length-$l$) of the sharp tip spike are further discussed in the following paragraphs.

\textbf{Effect of the spike diameter $(d)$: }

\begin{sidewaysfigure}
	\centering \includegraphics{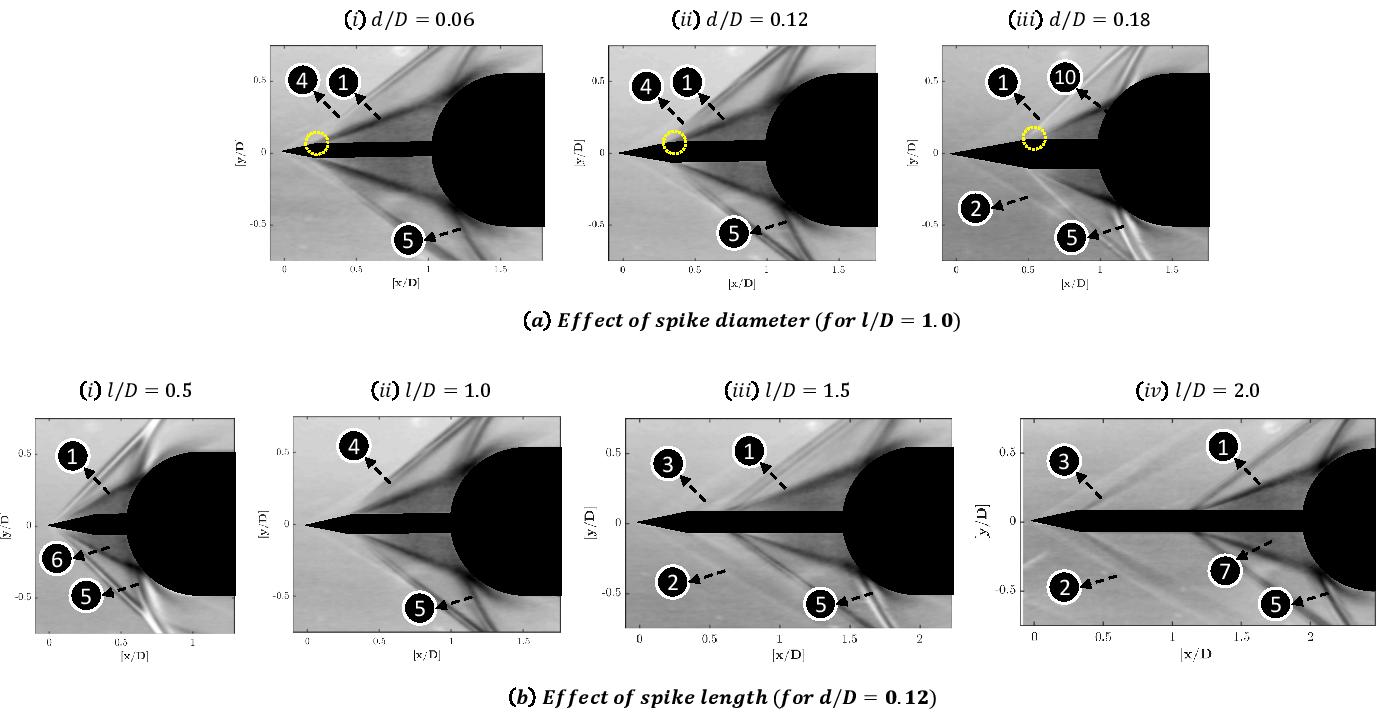}{} 
	\caption{Time-averaged shadowgraph image obtained through the operation of $\left\|\bar{\boldsymbol I}-\boldsymbol I_{rms}\right\|$ showing the (a) effect of spike diameter($d/D$), and (b) effect of spike length ($l/D$), on the dominant flow features around a hemispherical body at $M_\infty=2.0$. Dotted yellow circle shows the location of the separation point on the top surface as $d/D$ increases. Dominant flow features: 1. Separated shear layer, 2. Weak leading edge shock, 3. Expansion fan, 4. Separation shock, 5. Reattachment shock, 6. Re-circulation region (larger), 7. Re-circulation region (smaller), 8. Reattachment point, 9. Separation point. Flow is from left to right.}
	\label{figure9}
\end{sidewaysfigure}

\begin{figure}[htb!]
	\centering 
	\includegraphics{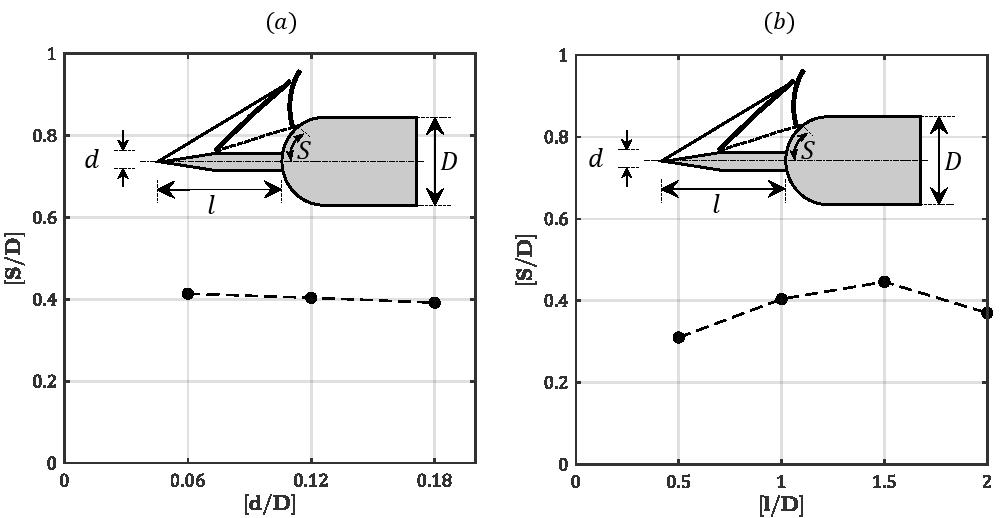}{}
	\caption{The effect of (a) spike diameter $(d/D)$ and (b) spike length $(l/D)$, on the location of the reattachment point formed over the hemisphere forebody at $M_\infty=2.0$. Obtained values are having an uncertainty of $\pm5\%$ as they are acquired from the time-averaged shadowgraph images. Solid circles represent the measured values. Dashed line shows the trend.}
	\label{figure10}
\end{figure}

In order to investigate the effect of the spike diameter ($ d $), sharp tip spikes of a constant length $(l/D=1)$ and different spike diameters $(d/D =0.06, 0.12 \; and \;0.18)$, have been used. The corresponding time-averaged shadowgraph images are shown in Figure~\ref{figure9}a. With the change in $d/D$, no significant change in the flow features is observed. The reattachment point is found to be generated around a fixed location for all the $d/D$. {However, a small variation in the size of the recirculation region can be noticed as the separation point moves radially outward (from $[y/D]=0.03$ to $[y/D]=0.09$) and approach towards the forebody (from $[x/D]=0.25$ to $[x/D]=0.5$) in \mbox{Figure \ref{figure9}a} (see the yellow dotted circles) with reference to the spatial location of the separation point for $d/D=0.06$ \mbox{(Figure \ref{figure9}a-(i))} as $d/D$ increases. The locations of the flow reattachment point ($ S/D $) have been obtained by performing a suitable spatial calibration on the time-averaged shadowgraph images for all the cases, and they are plotted in \mbox{Figure~\ref{figure10}}}.

{The closest point along the forebody surface where the reattachment shock and the imaginary line running along the center of the growing separated free shear layer intersect is considered as the reattachment point. An in-house Matlab routine is used to accomplish the procedure mentioned above}. In Figure~\ref{figure10}a, the variation in the location of the reattachment point with the increase in spike diameter ($d$) is shown. As stated, based on the shadowgraph images, no significant variation in the location of the reattachment point is observed. However, the flow turn angle near the reattachment point varies due to the increment of the cone angle formed by the re-circulation region at the point of separation. The separation point moves between {$[x/D]=0.25$ to $[x/D]=0.5$} as $d/D$ increases (see the yellow dotted circles in Figure \ref{figure9}a). Thus, with an increase in $d/D$, the cone angle increases for a constant reattachment point, and hence, the strength of the separation shock and flow turn angle near the reattachment point increases too. As a consequence, the strength of the reattachment shock will also increase. Hence, variations of similar order are expected in the wall static pressure near the reattachment point as well as in the $c_d$ values (will be discussed, eventually).

\begin{figure}[htb!]
	\centering \includegraphics[width=\linewidth]{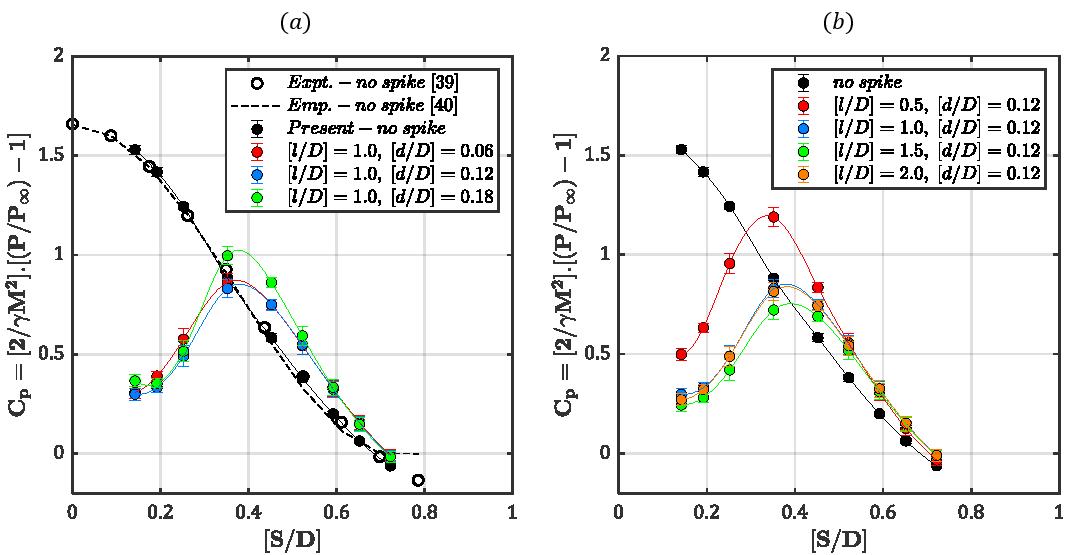}{}
	\caption{The effect of (a) spike diameter $(d/D)$ at $l/D=1.0$ and (b) spike length $(l/D)$ at $d/D=0.12$, on the coefficient of measured wall static pressure ($C_p = {2}(P-P_\infty)/{\gamma M^2 P_\infty}$) over the hemispherical forebody at $M_\infty=2.0$. Solid circles represent the measured values. Solid line shows the trend. Hollow circles show the experimental measurements on the hemispherical cylinder at $M_\infty=1.99$ and $\alpha=0^\circ$ from Ref.\protect \cite{Baer1961} and the dashed lines represent the empirical relation for the same conditions from Ref. \protect \cite{Perkins1952}.}
	\label{figure11}
\end{figure}

Surface or wall static pressure measurements have been carried out at different locations on the hemispherical forebody with and without a spike of different ($ d/D $), and the results are shown in Figure~\ref{figure11}a. Due to the spike, a significant reduction in the static pressure distribution is observed on the hemispherical forebody . Hence, a reduction in drag is also expected. However, as deduced from the analysis of the time-averaged shadowgraph images, a slight increase in static pressure distribution on the hemispherical forebody is observed while increasing the spike diameter from [$d/D$]=0.06 to [$d/D$]=0.18. The present findings, along with the postulates based on the time-averaged shadowgraph images, imply a slight increase in drag value with an increase in the spike diameter. The peak of each one of the pressure distribution curves obtained over the hemispherical spiked bodies corresponds to the location of the reattachment point on the hemispherical body from where the reattachment shock is found to be generated. On closely studying the pressure plots in Figure~\ref{figure11}a within the given spatial resolution in the sensor placements, it is observed that positions of the peaks of the pressure distributions are at the same location $\left(S/D\approx0.39\right)$ for all $d/D$ (see the solid trend line in Figure~\ref{figure11}a). This observation is consistent with the findings from the qualitative studies (using time-averaged shadowgraph images), stating that the reattachment point remains almost fixed while increasing $d/D$ (see Figure~\ref{figure10}a).

\begin{figure}[!htbp]
	\centering 
	\includegraphics{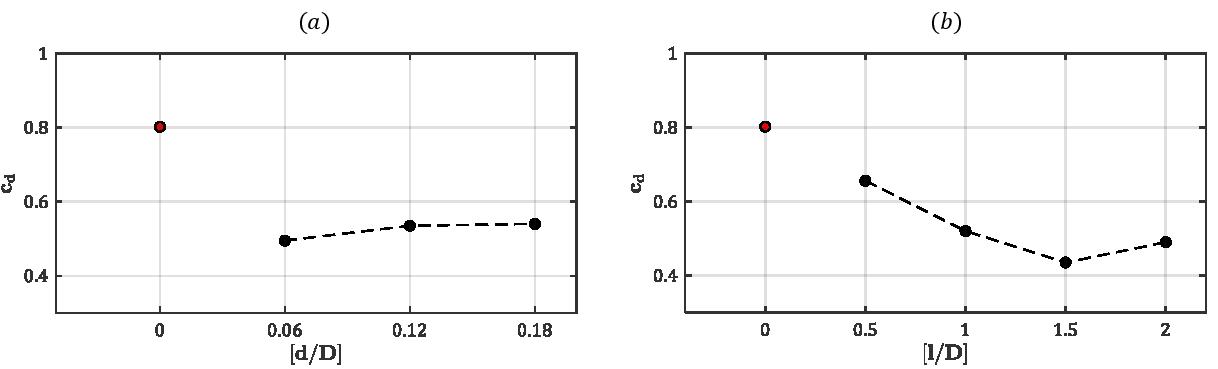}{} 
	\caption{The effect of (a) spike diameter $(d/D)$ at $l/D=1.0$ and (b) spike length $(l/D)$ at $d/D=0.12$, on the coefficient of measured forebody drag ($c_d$) at $M_\infty=2.0$. Black solid circles represent the experimental points and dashed lines are the trend lines. Red solid circle represents the measured coefficient of drag $(c_d)$ over a hemisphere without a spike ($l/D=0,\;d/D=0$).}
	\label{figure12}
\end{figure}

The forebody drag coefficient ($c_d$) for all the cases has also been measured using an in-house built strain gauge balance. The effect of $d/D$ on $c_d$ can be seen in Figure~\ref{figure12}a. As expected from the time-averaged shadowgraph images and static pressure measurements, a slight increase in the values of the drag coefficient with an increase of $d/D$ up to [$d/D$]=0.18 is observed. The measured $c_d$ are also tabulated in Table~\ref{table2}, indicating that within the uncertainty, a mild change in $c_d$ is observed with increase in $ d/D $.

\textbf{Effect of the spike length $(l)$: }

In order to investigate the effect of the spike length ($l/D$) on the time-averaged flow features, a sharp tip spike of constant stem diameter $d=0.12D$ has been used. The different spike lengths investigated in the present study are $l/D$=0.5, 1.0, 1.5, and 2.0. From Figure~\ref{figure9}b, it can be seen that, with an increase in $l/D$, the flow separation point along with the spike stem shifts downstream with a corresponding change in the separation shock strength as well. Such a shift in the separation point has been observed until a critical spike length of $l/D=1.5$. The re-circulation region between the hemispherical forebody and the spike (bounded by the separated free shear layer) also increases with an increase in $l/D$ up to $l/D=1.5$. On a further increase in $l/D>1.5$, no more downstream movement of the separation point has been observed from the hemispherical forebody. The change in the location of the separation point for increasing $l/D$, in turn, leads to the shifting of the reattachment point over the shoulder of the hemispherical forebody. The unlikely chance of reattachment point for different $l/D$ than for different $d/D$ can be attributed to the upstream flow conditions. As $l/D$ increases, the separation point occurs after the expansion fan near the sharp spike tip, especially for $l/D>1.0$. The separation point thus moves between $0-1.2D$ from the spike tip as $l/D$ increases. Such drastic variations alter the flow turning angle and strength of the separation and reattachment shocks. The thickening boundary layer on the spike stem as $l/D$ increases, further adds complexity, and results in moving the reattachment point upstream ($l/D=2.0$). The observations are consistent with the findings in Figure \ref{figure10}b. Similar results were also reported by Mair \unskip~\cite{279025:6283331}. 

The measured static pressure distribution for the blunt-body without and with a spike of different $l/D$ can be seen from Figure~\ref{figure11}b. {The results for the hemispherical forebody without spike case are compared with the previous experiments} \mbox{\cite{Baer1961}} {and empirical relations of Newton's sphere/hemisphere \mbox{\cite{Perkins1952}} in supersonic flow of similar $M_\infty$ and $\alpha$. Empirical values are slightly deviant near the expansion corner (forebody-cylinder interface) due to the inherent limitations in the matching of pressure and pressure gradients. Asides, the present values match well with the experiments and empirical relation, which add confidence to the current findings of hemispherical forebody mounted with spike}. Similar to Figure~\ref{figure11}a, a significant reduction in the time-averaged static pressure on the hemispherical forebody is observed due to the adoption of a spike. The solid trend lines (Figure~\ref{figure11}b) help in identifying the peaks of the pressure distribution curves for a spike mounted hemisphere. Each peak corresponds to the associated reattachment point, which moves downstream with an increase in $l/D$. This observation is consistent with those made from the time-averaged shadowgraph images (see Figure~\ref{figure9}b). The downstream movement of the reattachment point over the hemispherical forebody with an increase in $l/D$ takes place up to $l/D=1.5$, as seen in Figure~\ref{figure11}b (see the solid trend lines). A further increase in $l/D>1.5$ causes the reattachment point to move back to upstream resulting in a pressure distribution curve being similar to the one corresponding to the case of $l/D=1.0$. Thus, a reduction in the value of $c_d$ with increasing $l/D$ up to $l/D=1.5$ is anticipated. 

The effect of $l/D$ on $c_d$ obtained from the experiments is shown in Figure~\ref{figure12}b. In accordance with time-averaged shadowgraph and static pressure measurements, a reduction in the values of $c_d$ is observed with an increase $l/D$ up to $l/D=1.5$. A further increase in $l/D$ to $l/D=2.0$ results in increment of $c_d$ as expected from the discussions made in the previous paragraph. {Hence, it can be concluded that the percentage of $c_d$ reduction increase with increase in $l/D$ for a given $d/D=0.12$ up to an optimized spike length of $l/D=1.5$ in the present study.}

\subsubsection{Investigation of time-resolved flow field}The change in dominant flow features around the hemispherical forebody resulting from the mounting of a sharp tip spike has been observed and explained in the previous section. Considering the surface pressure distribution values and percentage of drag reduction achieved from the parametric time-averaged study, a sharp tip spike with [$l/D$]=1.5 shows a maximum drag reduction and stands out as the most efficient spike geometry. However, mounting of a drag-reducing spike also has a severe disadvantage associated with flow unsteadiness, which needs to be taken into consideration while looking for an efficient spike geometry. Therefore, a parametric time-resolved study is also required in order to compare the level of flow unsteadiness associated with the set of spike geometries included in the present study.

The existence of low-intensity shock-related unsteadiness in the case of a forward-facing step at a supersonic freestream Mach number has already been reported in \cite{david}. They reported that the free shear layer separating the re-circulation region from the external flow might become unstable (with respect to Kelvin-Helmholtz instabilities) and form large-scale structures. These structures interact with the separation shock and convect along the shear layer up to the reattachment point. They also reported that the charging and ejection of fluid mass in the re-circulation region is the driving mechanism for the shock-related unsteadiness. Thus, the shock-related unsteadiness associated with the hemispherical spiked body is expected to be similar with respect to the flow phenomena seen in shock-wave turbulent boundary layer interactions (SWTBL) problems as induced by compression ramps, reflected shocks, protrusions, and fins \unskip~\cite{279025:6296387}.

\begin{figure}[htb!]
	\centering 
	\includegraphics{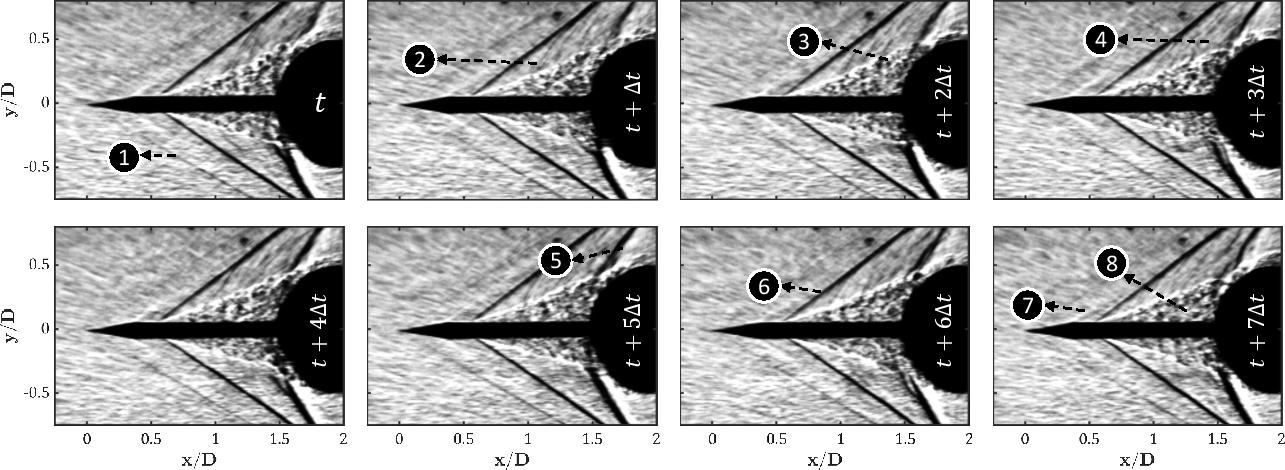}{}
	\caption{Instantaneous shadowgraph images obtained at different time intervals showing the difference in dominant flow features observed around a hemispherical forebody when mounted with a typical sharp tip spike ($l/D=1.5,\;d/D=0.12$) at $M_\infty=2.0$. Dominant flow features: 1. Leading edge shock, 2. Shocklet, 3. Large scale structures, 4. Shocklet Moving downstream, 5. Reattachment shock, 6. Separation shock, 7. Expansion fan, 8. Re-circulation region. Flow is from left to right.}
	\label{figure13}
\end{figure}

Figure~\ref{figure13} shows the time-resolved instantaneous shadowgraph images captured at time intervals of  $\Delta t=1/f $ (where, $f $ = 43000 Hz) over the hemispherical forebody mounted with a sharp tip spike of $l/D=1.5$ and $d/D=0.12$. In the supplementary, time-resolved shadowgraph video files corresponding to each of the cases are given.\footnote{Shadowgraph video filenames for different cases: $d/D=0.06$ in `video1', $d/D=0.12$ in `video2', $d/D=0.18$ in `video3', $l/D=0.5$ in `video4', $l/D=1.5$ in `video5', $l/D=2.0$ in `video6', and `no spike' in `video7'. Captions for each of the video files are given in a separate text file called `VideoCaptions.'} The time-resolved shadowgraph images indicate the generation of shocklets (as shown in Figure~\ref{figure4}) near the separation point and their downstream propagation along the free shear layer. With the passage of time, the generation of large-scale structures due to the existence of shear layer instabilities (Kelvin-Helmholtz type) can be observed convecting along the shear layer. The large-scale structures result in the formation of shocklets that move downstream along the shear layer and interact with the reattachment shock. Reattachment and separation shock foot oscillations associated with the charging and ejection of fluid mass from the re-circulation region are also observed in Figure~\ref{figure13}. For clarity, the reader is referred to see the video given in the supplementary under the name `video5' for the sharp tip spike case having a $l/D=1.5$, and $d/D=0.12$.

{In Figure \mbox{\ref{figure13}}, the separation points lying on the top and bottom surfaces of the spike stem are not symmetrical. Such an observation is even more prominent in the results shown later in the modal decomposition images (see Figure \mbox{\ref{figure17}}). There are two possible reasons for this asymmetry: a. variations in the angle of attack ($\alpha$) associated with the imperfection of the model mounting, and b. an asymmetrical oscillation or standing rotational wave \mbox{\cite{calarese,Demetriades}}. In the present case, given the uncertainty in positioning the model parallel to the freestream being less than $\pm$ $0.5^\circ$ (based on the accuracy of the inclinometer), the observation might be due to both the factors. The effects of $\alpha$ will alter the flow turning angle along the azimuth ($\phi$) and change the angle of the separated shear layer on the windward and leeward side. These effects are completely based on the spike tip shape, $l/D$, and $d/D$, which influence the net forces and moments acting on the body \mbox{\cite{reding,guenther_b,tahani}}. The gross flow features, which are discussed in the present paper, could be used to understand the dynamics involving variations in $\alpha$. Similarly, the changes observed on the intensity of the pressure fluctuations due to the standing rotational wave is a research topic by itself and hence beyond the scope of the current discussion which is restricted only to the dominant unsteadiness events arising due to the separated shear layer.}

\begin{figure}[htb!]
	\centering \includegraphics[width=0.5\textwidth]{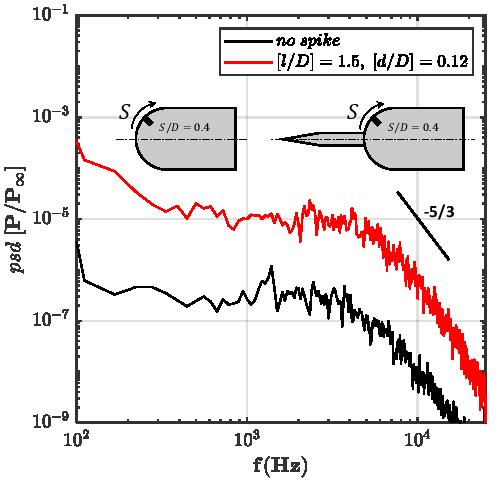}{}
	\caption{Power spectra of the measured pressure fluctuations near the shoulder $(S/D=0.4)$ of the hemispherical forebody configuration (line: black-clean hemisphere with no spike, red-hemisphere mounted with a sharp spike of $l/D=1.5$ and $d/D=0.12$) at $M_\infty=2.0$.}
	\label{figure14}
\end{figure}

To quantify the shock-related unsteadiness generated by mounting a sharp tip spike on a hemispherical body, unsteady pressure fluctuations have been measured near the shoulder ($ S/D=0.4 $) of the spiked forebody configurations as shown in Figure~\ref{figure7}. The power spectral density ($ psd $) of the pressure signal obtained from a hemisphere without (black line) and with a spike of {$l/D$=1.5 and $d/D$=0.12 (red line)} are compared and presented in Figure~\ref{figure14}. For the case of a hemispherical forebody mounted with a typical sharp tip spike, a broadband spectrum having a comparatively higher amplitude is observed, indicating a significant enhancement of flow unsteadiness. Similar studies have been conducted with variations in geometrical parameters (length, $l$ and diameter,$d$) of the sharp tip spike. 

In order to investigate the effect of the spike geometrical parameters on the shock related unsteadiness, sharp tip spike of different diameters ($d/D$=0.06 to 0.18, in steps of 0.06) and different lengths ($l/D$=0.5 to 2.0, in steps of 0.5) are considered. The level of pressure loading $\left(\zeta\right) $ and pressure fluctuations intensity $\left(\kappa\right) $ on the shoulder of the hemisphere have been derived from the unsteady pressure signals using equations \ref{eq-zeta} and \ref{eq-kappa}, respectively, and they are tabulated in Table~\ref{table2}. 
\begin{align}
\label{eq-zeta}
\zeta=\frac{P_{rms}}{P_\infty},\\
\label{eq-kappa}
\kappa=\frac{P_{s}}{P_{rms}}.
\end{align}
where,
\begin{equation*}
P_{rms}=\sqrt{\frac1n\sum_{i=1}^{n}P_i^{2}},\;P_{s}=\sqrt{\frac1n\sum_{i=1}^{n}({P_i-\overline P})^{2}}, P_i\;=\;\overline P+P',\;\overline P=\frac1n\sum_{i=1}^{n}P_i.
\end{equation*}

As seen in Table~\ref{table2}, with the increase in spike diameter ($d$), the values of the pressure loading $\left(\zeta\right) $ and associated $c_d$ increase. The reason for this is probably the increase in the strength of the reattachment shock. These findings are consistent with the postulates made regarding the reattachment shock strengths based on time-averaged shadowgraph images. A gradual reduction in the values of pressure loading $\left(\zeta\right) $ with increase in the spike length up to $l/D=1.5$ can also be seen in Table~\ref{table2}. Here again, the reason is probably the reduction in the strength of the reattachment shock. These findings are also consistent with the $c_d$ measurements and the postulates made from the time-averaged shadowgraph images as $l/D$ increases. Furthermore, the pressure fluctuation intensity $\left(\kappa\right) $ is reduced with an increase in the spike diameter. It can be observed from Table~\ref{table2} that the value of $\kappa $ increases with an increase in the spike length up to $l/D=1.5$ and remains almost constant as $l/D$ increases.

\begin{table}[hbt!]
	\caption{ Details of the measured drag coefficient ($c_d$), calculated pressure loading ($\zeta$), and pressure fluctuation intensity ($\kappa$) in the hemisphere mounted with different spike configurations at $M_\infty=2.0$.}
	\label{table2}
	\centering 
	\begin{tabular}{cccccccc}
		\hline
		\multirow{2}{*}{Spike} & \multirow{2}{*}{No} & \multicolumn{3}{c}{$l/D=1.0$} & \multicolumn{3}{c}{$d/D=0.12$} \\ \cline{3-8} 
		Configuration & Spike & $d/D=0.06$ & $d/D=0.12$ & $d/D=0.18$ & $l/D=0.5$ & $l/D=1.5$ & $l/D=2.0$ \\ \hline
		\begin{tabular}[c]{@{}c@{}}$c_d(\pm 5\%)$\\ (Drag Coefficient)\end{tabular} & 0.8 & 0.5 & 0.52 & 0.54 & 0.67 & 0.43 & 0.49  \\
		\begin{tabular}[c]{@{}c@{}}$\zeta(\pm 5\%)$\\ (Pressure Loading)\end{tabular} & 2.89 & 2.8 & 3.2 & 3.5 & 4.1 & 2.6 & 3 \\
		\begin{tabular}[c]{@{}c@{}}$\kappa(\pm 5\%)$ (Pressure\\ Fluctuation Intensity, $\%$)\end{tabular} & 1.6 & 12 & 11 & 8 & 8 & 14 & 13 \\ \hline
	\end{tabular} 
\end{table}

\textbf{Effect of the spike diameter $(d)$: }

\begin{sidewaysfigure}
	\centering \includegraphics{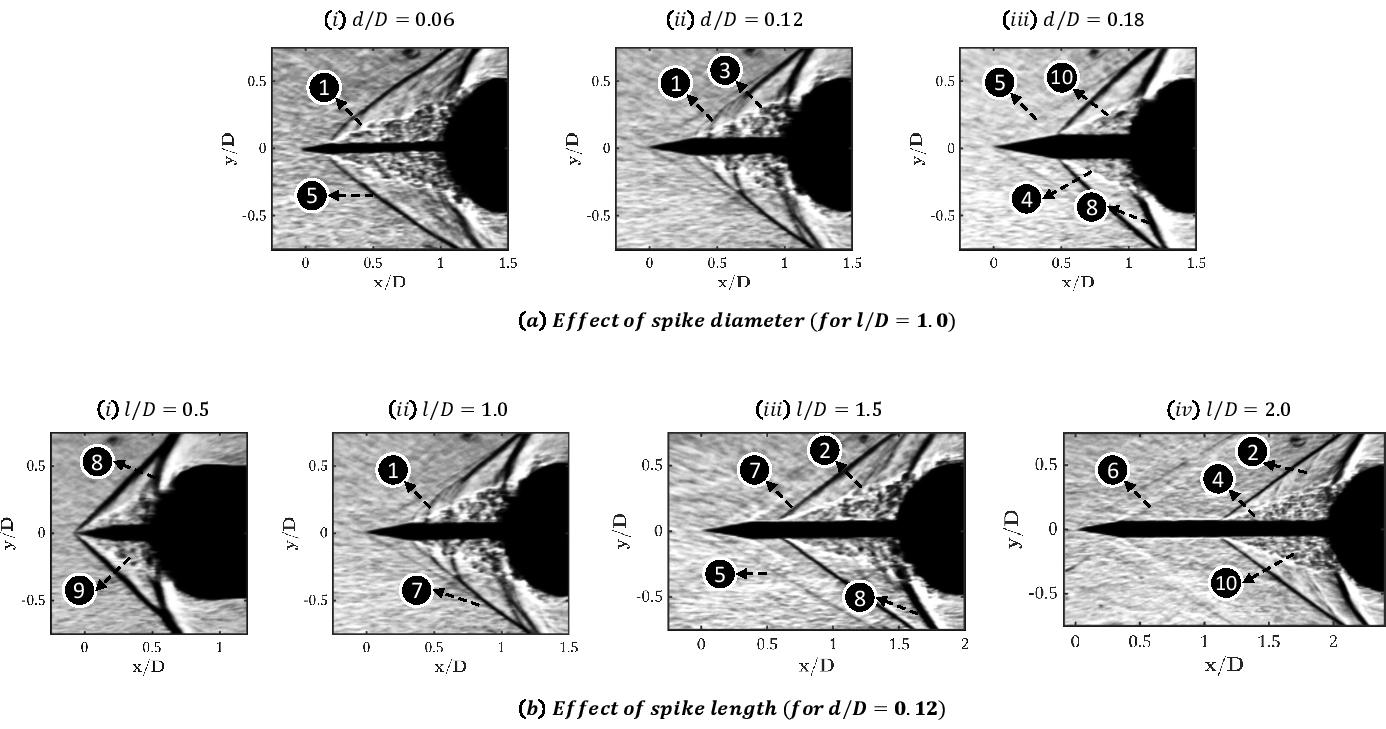}{} 
	\caption{{Instantaneous shadowgraph images showing the effect of (a) spike diameter ($d/D$) and (b) spike length ($l/D$), on the dominant flow features around a hemispherical forebody at $M_\infty=2.0$. Dominant flow features: 1. Shocklets (stronger), 2. Shocklets (weaker), 3. Large scale structures, 4. Finer scale structures, 5. Weak leading edge shock, 6. Expansion fan, 7. Separation shock, 8. Reattachment shock, 9. re-circulation region (larger), 10. re-circulation region (smaller), 11. Separated shear layer. Flow is from left to right.}}
	\label{figure15}
\end{sidewaysfigure}

The instantaneous frames of time-resolved shadowgraph images captured over a hemisphere mounted with a sharp tip spike of different diameters ($d$) and lengths ($l$) are presented in Figure~\ref{figure15}a and b. From Figure~\ref{figure15}a, it can be seen that with the increase in $d/D$, the size of the large-scale structures (see Figure~\ref{figure4}) formed along the free shear layer is reduced. This, in turn, reduces the strength of the convecting shocklets (as shown in Figure~\ref{figure4}) formed along the free shear layer resulting in a lesser oscillation of the reattachment shock. 

\begin{figure}[htb!]
	\centering \includegraphics{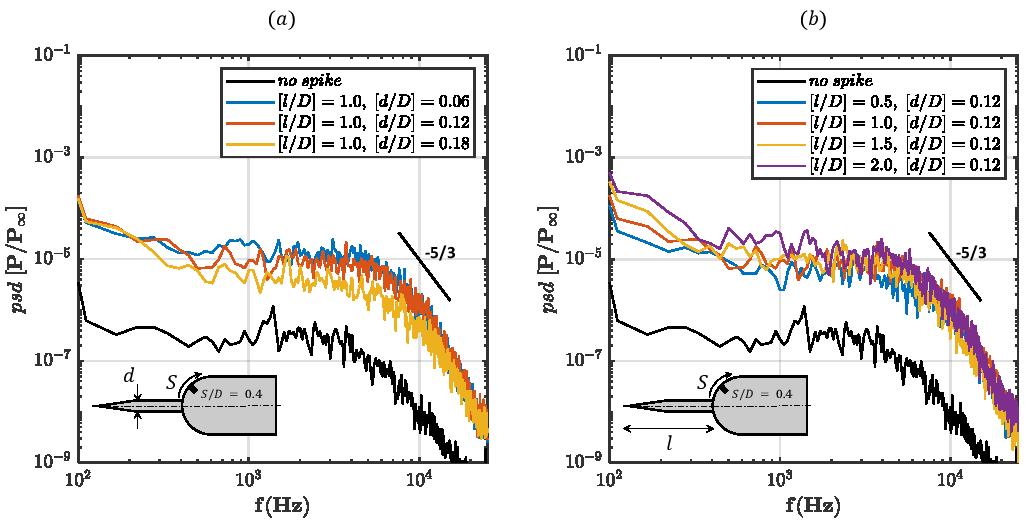}{} 
	\caption{Power spectra of the measured pressure fluctuations near the shoulder $(S/D=0.4)$ of the hemispherical forebody configuration showing the effect of (a) spike diameter ($d/D$) and (b) spike length ($l/D$), at $M_\infty=2.0$.}
	\label{figure16}
\end{figure}

\begin{sidewaysfigure}
	\centering \includegraphics{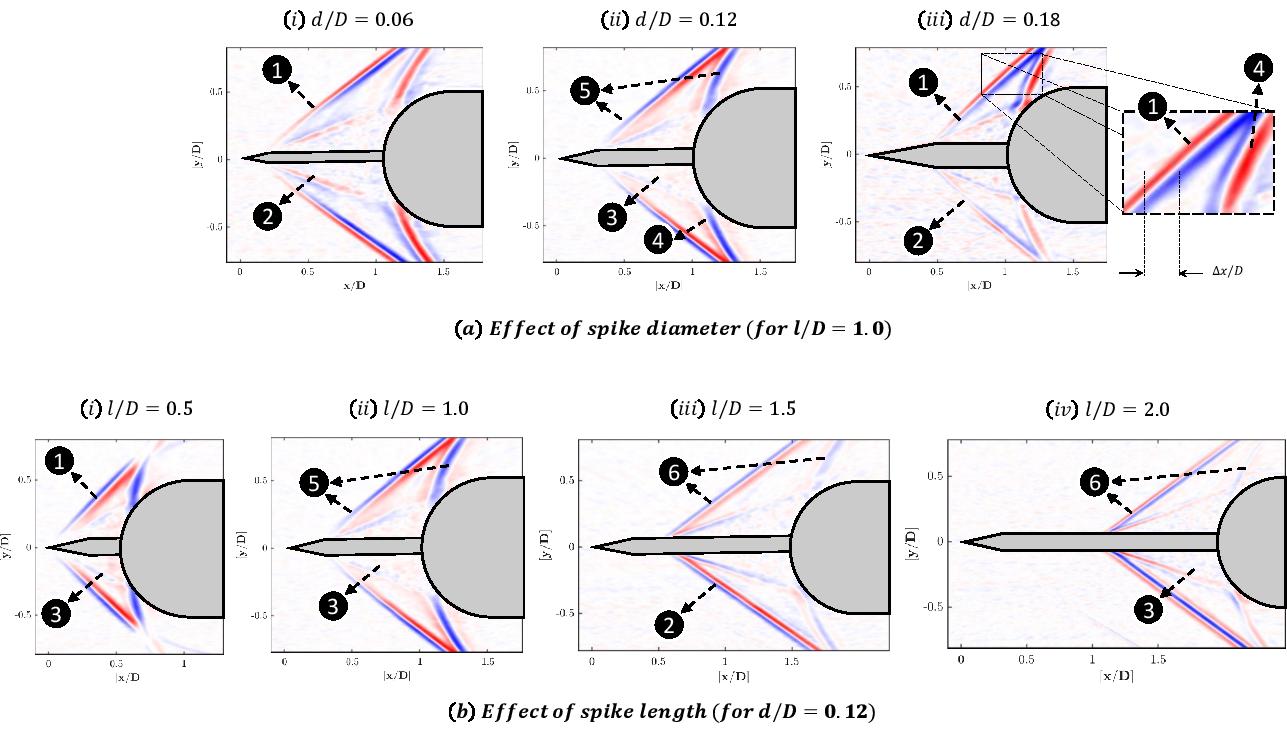}{} 
	\caption{Dominant energetic spatial mode $\left[\Phi_1(x,y)\right] $ obtained from the POD analysis of shadowgraph images taken from the hemispherical spiked body configuration mounted with a sharp tip spike of (a) different diameters ($d/D$) for a $l/D=1.0$, and (b) different lengths ($l/D$) for a $d/D=0.12$ at $M_\infty=2.0$. Flow features: 1. Unsteady separation shock, 2. Separated shear layer, 3. Re-circulation region, 4. Reattachment shock, 5. Strong coupling between separated and reattached shock oscillations, 6. Weak coupling between separated and reattached shock oscillations. Flow is from left to right.}
	\label{figure17}
\end{sidewaysfigure}

Spectral analysis has been made using the measured pressure fluctuations at the location near the shoulder of the hemisphere ($ S/D=0.4 $, see Figure~\ref{figure7}) for all cases considered. The effect of change in spike diameter ($d$) on the power spectra are presented in Figure~\ref{figure16}a. Broadband spectra for all the cases are exhibited with higher amplitudes compared to the case with no spike. In particular, the range of frequencies between 2-8 kHz seems to be amplified and is probably associated with the amplification of shear layer instabilities. The difference in the spectra with a change in the $d/D$ is noticeable at a range of frequencies around 5 kHz. As also observed from the time-resolved instantaneous shadowgraph frames, the intensity of the shock-related unsteadiness is reduced with an increase in $d/D$. The highest/lowest intensity of the power spectra is observed for the spike with a minimum/maximum diameter of $d/D=0.06/0.18$, respectively (blue/yellow line of Figure~\ref{figure16}a).

\textbf{Effect of the spike length $(l)$: }

Similar studies have been conducted in order to investigate the effect of the spike length ($l $) on the time-resolved flow features around the hemispherical spiked body configuration. It is observed that the size of the large-scale structures formed along the free shear layer is slightly increased with an increase in the spike length up to $l/D=1.5$, as shown in Figure~\ref{figure15}b. A further increase in the spike length to $l/D=2.0$ does not affect the size of the large-scale structures. With an increase in $l/D$, the flow separation point on the spike stem moves downstream along the spike and thereby increasing the separation shock angle. As mentioned in the previous section, this leads to a lower reduction in the Mach number behind the separation shock. Consequently, the convective Mach number across the shear layer is increased, as reported by Slessor et al., \unskip~\cite{279025:6948594}. This leads to a lesser growth rate of the shear layer. Lower growth rate reduces the size of the large scale structures and thus the observation of increasing pressure fluctuations intensity $(\kappa)$ as shown in Table~\ref{table2}. A similar trend but to a lesser extent is observed in the power spectrum {\mbox{(Figure \ref{figure16}b)}} over the hemisphere mounted with the sharp tip spike of different $l/D$. {As explained earlier, with the increase in the spike diameter ($d$), the flow separation point moves downstream (from $[x/D]=0.25$ to $[x/D]=0.5$) resulting in the reduction of the length of the shear layer.} This, in turn, reduces the growth of the large-scale structure formed along the shear layer. In addition, an increase in $d$, further reduces the angle formed by the re-circulation region at the point of separation, thereby reducing the flow turning angle required ahead of the reattachment point. Thus the power spectrum (Figure \ref{figure16}a) from the unsteady pressure measurements shows a decreasing trend for increasing $d$. 

The effect of the spike diameter ($d$) and spike length $(l)$ on the shock-related unsteadiness discussed so far is based upon the results obtained from the time-resolved instantaneous shadowgraph images and the unsteady pressure measurements. Furthermore, an attempt has been made to verify our findings using the modal analysis of the time-resolved shadowgraph images. For that purpose, both POD and DMD analysis have been carried out, and the dominant energetic spatial mode ($\Phi_1\left(x,y\right)$) and dynamic temporal mode ($\alpha\left[\Theta\left(x,y\right)\right],f$) have been computed for the case of a hemispherical forebody mounted with different spikes. The dominant energetic spatial mode $\left[\Phi_1\left(x,y\right)\right] $ represents the corresponding time-averaged flow field image as shown in Figure~\ref{figure17}. Only time-varying dominant flow features are observed in Figure \ref{figure17}. The absence of leading-edge shock in the dominant energetic modes verifies the fact that it is steady. The horizontal distance between the extrema of the colour contours near the separation and reattachment shocks in Figure~\ref{figure17} marks the extrema of shock oscillations. It is observed that the phenomenon of separation and reattachment shocks oscillation exist in all the investigated cases of spike mounted hemispherical forebody configurations. However, the magnitudes of the separation and the reattachment shocks oscillation (Figure \ref{figure17}a) are found to be of similar magnitude with the change in the spike diameter ($d$) with $\Delta x/D=0.15$ (separation shock) and $\Delta x/D=0.20$ (reattachment shock). In addition, the existence of opposite band of colour contour (for example in Figure \ref{figure17}a-iii, the separation shock has blue-red contour band and the reattachment shock has red-blue contour band, but of similar magnitude in $\Delta x/D$), means that the oscillations of the separation and reattachment shocks are strongly coupled and have negative correlation \cite{kutz} (or out-of-phase shock motion). Such kind of observation is consistent with the conclusions provided in \cite{david} in regards to charging and ejection of fluid mass in the re-circulation region for the forward-facing step in a supersonic stream, where a distinct out-of-phase movement of the separated and reattachment shocks are seen. In case of varying spike length $(l)$ as shown in Figure~\ref{figure17}b, the shock oscillation intensity is observed to be decreasing and the coupling between the separation and reattachment shocks weakens. These observations corroborate with the shadowgraph images shown in Figure~\ref{figure9} and Figure~\ref{figure15}.

\begin{figure}[htb!]
	\centering \includegraphics{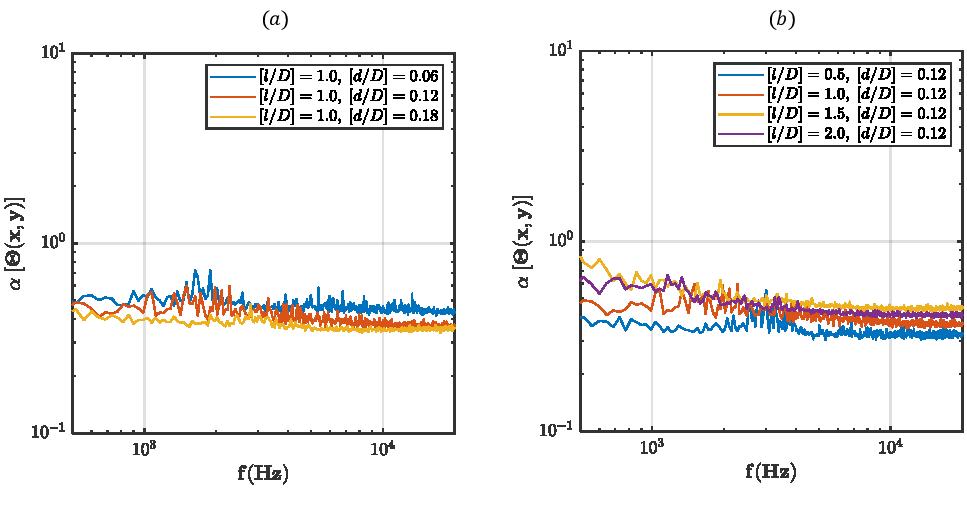}{}
	\caption{The dynamic spectra observed from the DMD ($\Theta(x,y)$) analysis for the hemisphere mounted with a sharp tip spike of (a) different spike diameters ($d/D$) for $l/D=1.0$ and (b) different spike lengths ($l/D$) for $d/D=0.12$ at $M_\infty=2.0$.}
	\label{figure18}
\end{figure}

The dynamic spectra have been obtained from DMD for all the considered cases, and they are compared in Figure~\ref{figure18}. The findings from the DMD analysis are consistent with the spectral contents obtained from the pressure measurements (see Figure~\ref{figure16}). From the analysis of Figure~\ref{figure18}, it can be seen that the unsteadiness associated with the hemispherical spiked body configuration reduces with an increase in $d/D$ (see Figure~\ref{figure18}a). The sharp tip spike having the maximum $d/D$ has the least shock-related unsteadiness. On the other hand, the intensity of the shock-related unsteadiness is increased with an increase in the spike length up to $l/D=1.5$, as shown in Figure~\ref{figure18}b. A further increase in $l/D$ to $l/D=2.0$ does not affect the intensity of the shock-related unsteadiness and remains almost the same as in the case of $l/D=1.5$. Hence, our findings from the modal analysis support the postulates made from the time-resolved shadowgraph images and unsteady pressure measurements.

\subsection{Effect of spike tip geometry}
\begin{figure}[]
	\centering \includegraphics[width=0.7\textwidth]{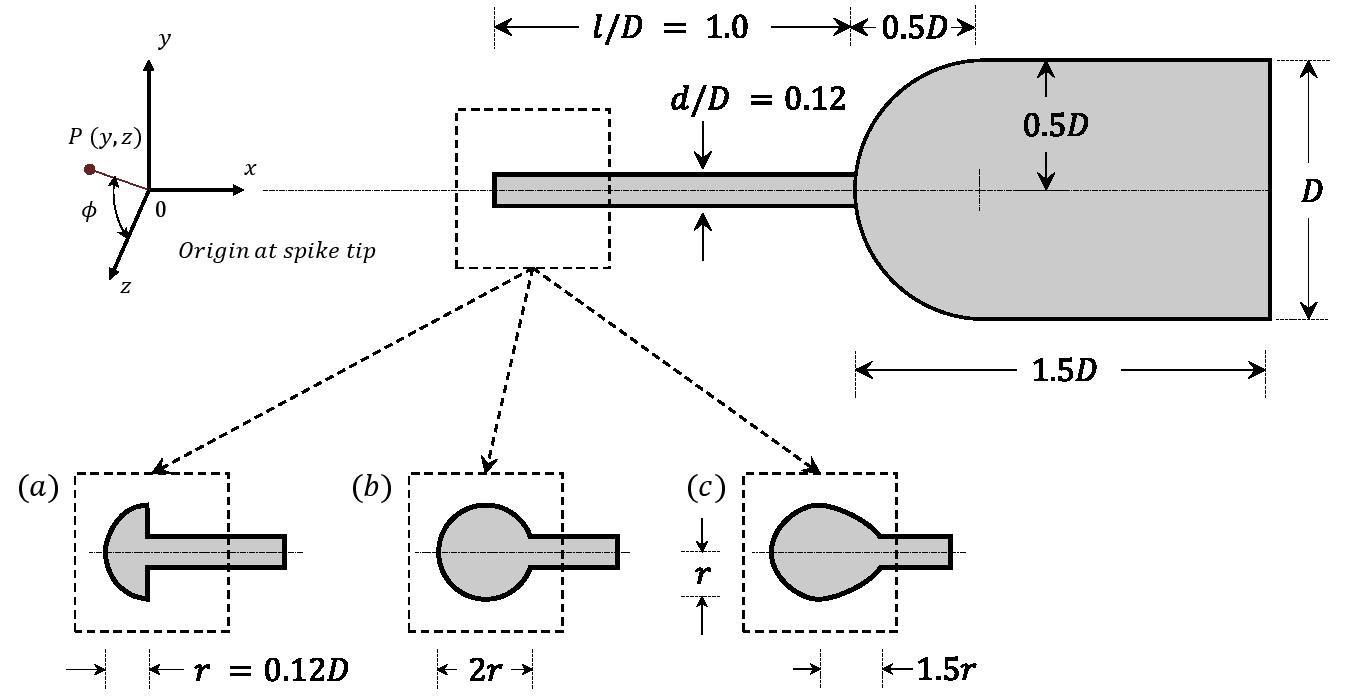}{}
	\caption{Schematic (not drawn to scale) showing the primary geometrical parameters of the hemispherical tip spike on a hemispherical spiked body configuration. Different base shapes of the hemispherical spike tip are shown: (a). Vertical base, (b). Circular base, and (c). Elliptical base. The origin is at the spike tip. Flow is from left to right.}
	\label{figure19}
\end{figure}

In the previous section, we have shown the effect of the diameter ($d$) and the length ($l$) of a sharp tip spike on the time-averaged and time-resolved flow features generated around the hemispherical spiked body. From the parametric studies conducted so far, the sharp tip spike with $l/D=1.5$ and $d/D=0.12$ is found to be the most effective one in terms of drag reduction (see Figure~\ref{figure12}b), but it generates a considerable amount of shock-related unsteadiness (see Figure~\ref{figure16}b-yellow line). Similarly, the sharp tip spike with $l/D=1.0$ and $d/D=0.18$ generates the minimal amount of shock-related unsteadiness (see Figure~\ref{figure16}a-yellow line) but the drag reduction in this case is quite low (see Figure~\ref{figure12}a). In order to devise a compromised spike geometry having a better drag reduction while generating a minimal level of shock-related unsteadiness, the tip of the sharp spike has been altered. Based on the investigations of the shock-related unsteadiness over the hemispherical spiked body configuration and the parameters influencing the level of unsteadiness, a suitable spike tip of hemispherical shape with different base shapes (vertical, circular, and elliptical) has been adopted. Hence, further experiments have been conducted on a hemispherical spiked body configuration with a spike of a hemispherical tip having its radius being equal to the diameter of the spike stem ($d$). The overall length of the spike is kept equal to the forebody diameter ($l/D$=1.0). The effect of the spike length and the spike stem diameter with a hemispherical spike tip has not been investigated in the present study. The geometrical details of the hemispherical spike tip with different base shapes are shown in Figure~\ref{figure19}. 

\begin{table}[hbt!]
	\caption{ Details of the measured drag coefficient ($c_d$) and the calculated pressure loading ($\zeta$) and pressure fluctuation intensity ($\kappa$) in the hemispherical forebody mounted without and with spike tip of different shapes $(l/D=1.0,\;d/D=0.12)$ at $M_\infty=2.0$.}
	\label{table3}
	\centering 
	\begin{tabular}{cccc}
		\hline
		Spike Configurations    & $c_d (\pm 5\%)$ & $\zeta (\pm 5\%)$ & $\kappa (\pm 5\%)$, $\%$ \\
		\hline
		Without spike (hemispherical forebody) & 0.8 & 2.89 & 1.6 \\    
		Sharp spike tip ($l/D=1.0,\;d/D=0.12$) & 0.52 & 3.2 & 11 \\
		Hemispherical spike tip (vertical base) & 0.36 & 2.04 & 10 \\
		Hemispherical spike tip (circular base) & 0.38 & 2.11 & 12 \\
		Hemispherical spike tip (elliptical base) & 0.39 & 2.22 & 11\\
		\hline 
	\end{tabular} 
\end{table}

\subsubsection{Investigation of time-averaged flow field}
\begin{figure}[htb!]
	\centering \includegraphics[width=\textwidth]{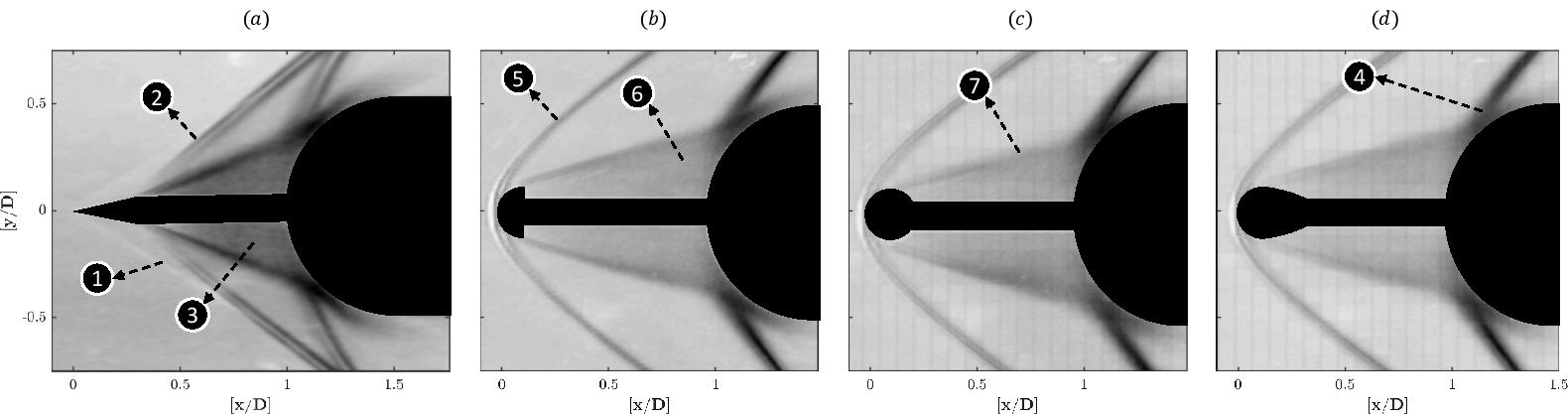}{}
	\caption{Time-averaged shadowgraph image obtained through the operation of $\left\|\bar{\boldsymbol I}-\boldsymbol I_{rms}\right\|$ for the hemispherical forebody mounted with (a) a sharp tip spike ($l/D=1.0,\;d/D=0.12$), (b) a hemispherical spike tip with a vertical base ($l/D=1.0,\;d/D=0.12$), (c) a hemispherical spike tip with a circular base ($l/D=1.0,\;d/D=0.12$), and (d) a hemispherical spike tip with an elliptical base ($l/D=1.0,\;d/D=0.12$) at $M_\infty=2.0$. Dominant flow features: 1. Attached leading edge shock, 2. Separation shock, 3. Re-circulation region (smaller), 4. Reattachment shock, 5. Detached bow shock, 6. Re-circulation region (larger), 7. Separated free shear layer. Flow is from left to right.}
	\label{figure20}
\end{figure}

\begin{figure}[htb!]
	\centering \includegraphics[width=0.8\linewidth]{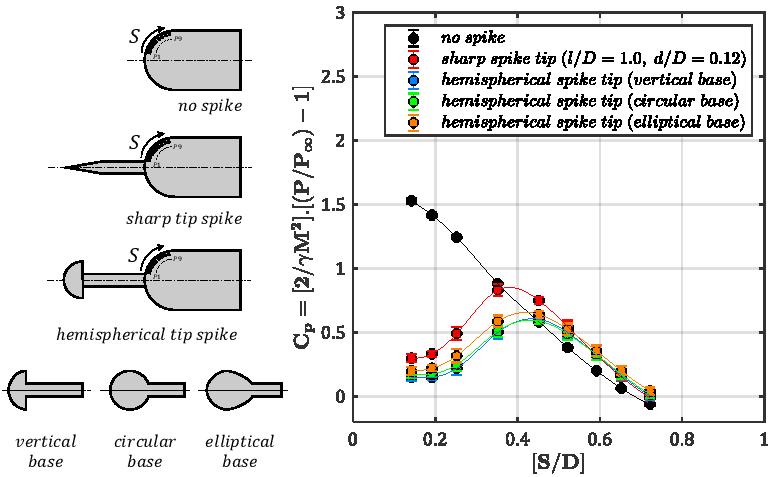}{}
	\caption{The effect of spike tip on the coefficient of the measured wall static pressure ($C_p = {2}(P-P_\infty)/{\gamma M^2 P_\infty}$) over the hemispherical forebody without and with spike of different spike-tip shapes ($l/D=1.0,\;d/D=0.12$) at $M_\infty=2.0$. Solid line shows the trend.}
	\label{figure21}
\end{figure}

The time-averaged images (through the operation of $\left\|\bar{\boldsymbol I}-\boldsymbol I_{rms}\right\|$) obtained from set of 1000 shadowgraph images captured at a high frame rate of 43000 Hz and exposure time of 2 $\mu s$ are shown in Figure~\ref{figure20}. The formation of a detached weak bow shock ahead of the hemispherical spike tip (see Figure~\ref{figure20}b-d) unlike the attached separation shock formed from the spike body (see Figure~\ref{figure20}a) in the case of a sharp tip spike is observed. From the time-averaged shadowgraph images, it is noticed that the intensity of the reattachment shock is reduced by changing the shape of the spike tip from a sharp tip to a hemispherical tip. This would in turn lead to a reduction in the level of the surface pressure distribution over the hemisphere and thereby reducing the associated forebody drag. The surface pressure distribution measured over the hemispherical body mounted with a hemispherical spike tip is shown in Figure~\ref{figure21}. A reduction in the surface pressure level to a considerable extent is observed for the case of hemispherical spike tip in comparison to the case of a sharp tip spike. The peak of the pressure plot corresponding to the reattachment point is seen to move downstream (see the solid trend lines in Figure~\ref{figure21}). One of the primary reasons is due to the increase in the size of the recirculation region associated with the vertical movement of the separated free shear layer caused by changing the spike tip geometry. The separated free shear layer from the hemispherical spike tip now forms a smaller separation angle with the body axis so that the flow leaves the forebody tangentially and hence the downstream movement of the reattachment point. A reduction in the flow turn angle near the shoulder of the hemispherical forebody due to the downstream movement of the reattachment point results in lowering the strength of the reattachment shock and a reduction in the associated level of the pressure distribution (see Table \ref{table3}). In addition, the formation of a detached bow shock wave ahead of the hemispherical spike tip reduces the Mach number to a greater extent and again weakens the strength of the reattachment shock formed further downstream near the hemisphere forebody. The reduction in the surface pressure distribution and an increase in the size of the recirculation region should result in a further reduction in $c_d$. As tabulated in Table \ref{table3}, a reduction in $c_d$ is indeed observed by 55\% and 35\% in comparison with the forebody having no spike and with a sharp tip spike ($l/D=1.0,\;d/D=0.12$). As the spike base changes from vertical to circular and elliptical, the gross flow features remain the same. However, from the static pressure measurements (Figure \ref{figure21}), elliptical base shows a slight increment suggesting a little higher reattachment shock strength. The values of $c_d$ is also little larger (Table \ref{table3}), primarily due to the fact of slightly higher $\zeta$ but also due to the fact that the base surface area is considerably increased.

\subsubsection{Investigation of time-resolved studies}
\begin{figure}[htb!]
	\centering 
	\includegraphics{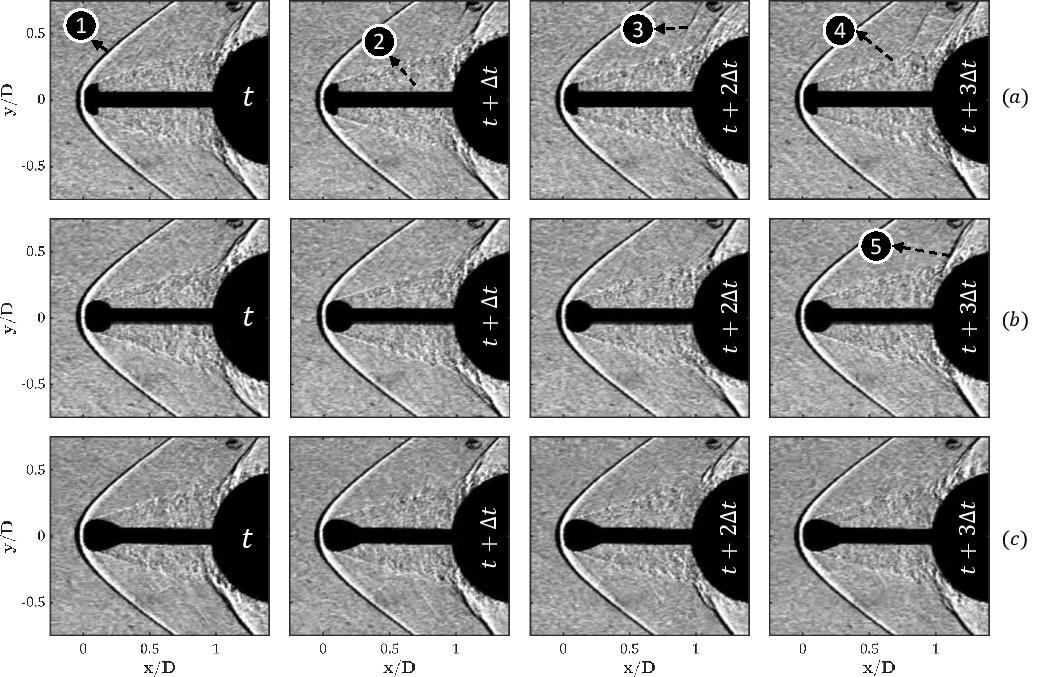}{}
	\caption{Instantaneous shadowgraph images at different time intervals around a hemispherical forebody when mounted with a hemispherical tip spike ($l/D=1.0,\;d/D=0.12$) of different base shapes: (a) vertical, (b) circular, and (c) elliptical at $M_\infty=2.0$. Dominant flow features: 1. Detached bow shock, 2. Re-circulation region, 3. Weak shocklets, 4. Large-scale structures in the separated free shear layer, 5. Reattachment shock. Flow is from left to right.}
	\label{figure22}
\end{figure}

\begin{figure}[htb!]
	\centering 
	\includegraphics[width=0.8\textwidth]{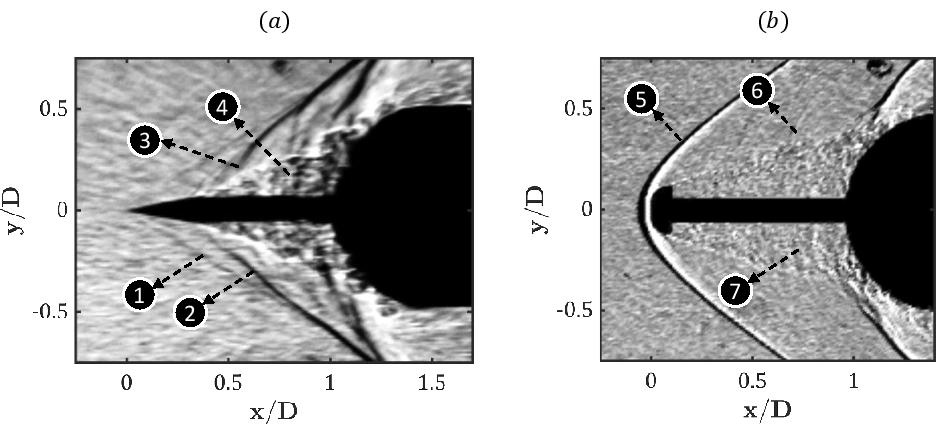}{}
	\caption{Instantaneous shadowgraph images showing the comparison of the dominant flow features observed around a hemispherical body mounted with (a) a sharp tip spike ($l/D=1.0,\;d/D=0.12$) and (b) a hemispherical tip spike ($l/D=1.0,\;d/D=0.12$), at a freestream supersonic flow Mach number of $M_\infty=2.0$. Dominant flow features: 1. Attached leading edge shock, 2. Separation shock, 3. Shocklets (stronger), 4. re-circulation region (smaller), 5. Detached bow shock, 6. Shocklets (weaker), 7. re-circulation region (larger). Flow is from left to right.}
	\label{figure23}
\end{figure}

In order to investigate the effect of changing the geometry of spike tip on the shock-related unsteadiness, time-resolved studies have been conducted. The instantaneous shadowgraph images\footnote{Corresponding time-resolved shadowgraph video file is given in the supplementary under the name `video8', `video9', and `video10'.} captured at different time intervals over a hemispherical body mounted with a spike having a hemispherical tip of different shapes are shown in Figure~\ref{figure22}. It can be observed that the generation of a detached shock and the location of the flow separation point just behind the shoulder of the hemispherical tip, cause the elimination of the separation shock and thereby the shock-wave turbulent boundary layer interactions (SWTBL) are restricted. {The primary reason behind the elimination of the separation shock is due to the fact that the flow behind the bow shock accelerates along the surface of the hemispherical spike tip with a negligible boundary layer thickness owing to the favorable pressure gradient.} Consequently, only weak shocklets are formed and are barely visible in Figure~\ref{figure22}a. For different base shapes, the gross flow features remain the same. Furthermore, the large-scale structures formed along the free shear layer are observed to be smaller in size for the case of a hemispherical spike tip when compared to the structure associated with the sharp tip spike of the same length. A comparison of instantaneous time-resolved shadowgraph images\footnote{Corresponding time-resolved shadowgraph video file is given in the supplementary under the name `video2', and `video8'.} captured over a hemispherical body mounted with a sharp tip spike and a hemispherical tip spike is presented in Figure~\ref{figure23}. The presence of weaker shocklets and the reduction in the size of the large-scale structures by changing the spike tip geometry from a sharp tip to a hemispherical tip means that a reduction in the intensity of the shock-related unsteadiness is possible. 

\begin{figure}[htb!]
	\centering \includegraphics[width=0.8\textwidth]{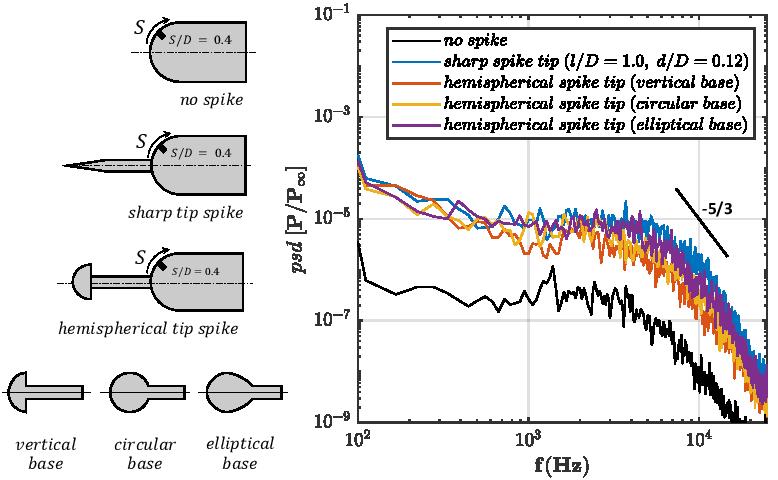}{}
	\caption{Power spectral densities measured near the shoulder $(S/D=0.4)$ of the hemispherical forebody configuration without and with spike of different spike-tip shapes ($l/D=1.0,\;d/D=0.12$) at $M_\infty=2.0$.}
	\label{figure24}
\end{figure}

To further support the above-mentioned possibilities, unsteady pressure fluctuations have been measured at the location near the hemisphere shoulder ($S/D=0.4$) for the hemispherical body mounted with the hemispherical spike tip. As Tabulated in Table \ref{table3}, a reduction of 39\% in the value of  $\zeta $ is observed. Similarly, the pressure fluctuation intensity $\left(\kappa\right) $ is also reduced to 10\% for the case of the hemispherical tip spike, which is a bit lower compared to the case of the sharp tip spike of same $l/D$ and $d/D$. The pressure spectra obtained from the measured pressure fluctuations are compared in Figure~\ref{figure24} with those obtained for the case of a sharp tip spike of $l/D=1.0$ and $d/D=0.12$. It can be clearly seen that the intensity of the shock-related unsteadiness is reduced by changing the spike tip geometry to a hemispherical tip from a sharp tip spike. While changing the base shapes of the hemispherical spike tip, smaller variations are observed. The communication of wave-fronts/disturbances from the forebody to the separated free shear layer prevents the movement of separation point in case of the vertical base, and it is not in the case of the circular and elliptical base. {The abrupt flow turn angle and the presence of a favorable pressure gradient along with the spike tip aid in preventing the point of separation from moving back in the vertical base. However, in the case of the hemispherical and elliptical base, the flow turning is gradual, and the separation point moves slightly downstream and gains a degree of freedom to accommodate for the disturbances by weakly moving forth and back, just like in the case of the separation point for a sphere in a supersonic flow \mbox{\cite{nagata}}. This results in continuous communication of weak disturbances along the separated free shear layer and hence a moderate increase in $\kappa$ \mbox{(Table \ref{table3})}. Such a movement also changes the re-circulation region size, reattachment point location, and slightly higher $\zeta$ \mbox{(Table \ref{table3})} for the circular and elliptical base. Observations from \mbox{Figure \ref{figure24}} is also in accordance with the above statements.} 

\begin{figure}[htb!]
	\centering \includegraphics[width=\textwidth]{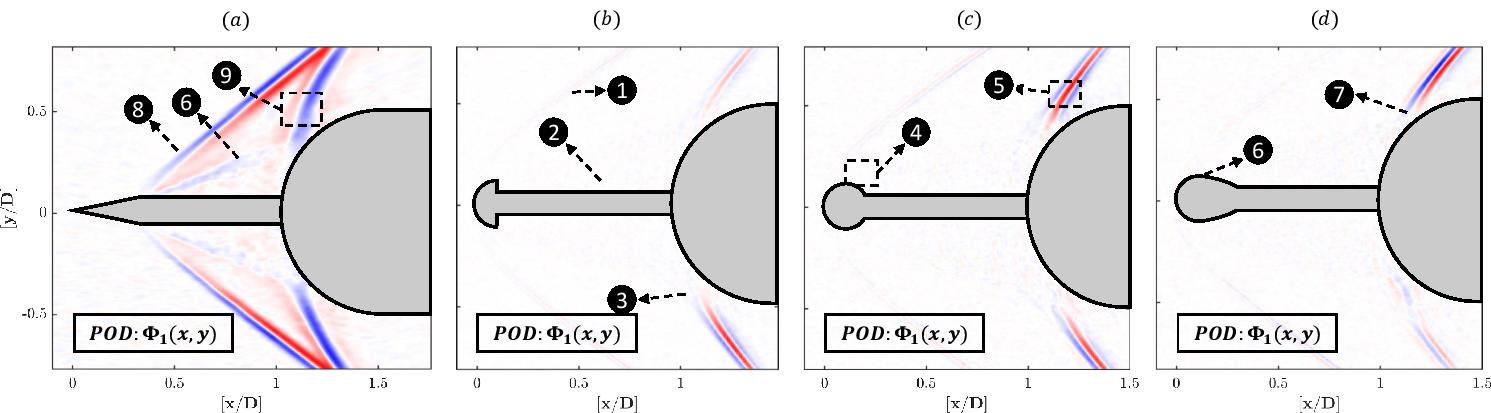}{}
	\caption{Dominant energetic spatial mode $\left[\Phi_1\left(x,y\right)\right]$ obtained from the POD analysis of shadowgraph images of the hemispherical body configuration mounted with (a) a typical sharp tip spike (b) a hemispherical spike tip with a vertical base ($l/D=1.0,\;d/D=0.12$), (c) a hemispherical spike tip with a circular base ($l/D=1.0,\;d/D=0.12$), and (d) a hemispherical spike tip with an elliptical base ($l/D=1.0,\;d/D=0.12$), at $M_\infty=2.0$. Flow features: 1. Steady/weakly moving detached bow shock, 2. Re-circulation region, 3. Separated free shear layer and reattachment shock interaction zone, 4. Absence of separated shock motion and mildly unsteady separated shear layer, 5. Oscillating reattachment shock (weaker), 6. Unsteady separated shear layer, 7. Unsteady reattachment shock, 8. Moving separated shock, 9. Moving reattachment shock (stronger). Flow is from left to right.}
	\label{figure25}
\end{figure}

\begin{figure}[htb!]
	\centering \includegraphics[width=0.51\textwidth]{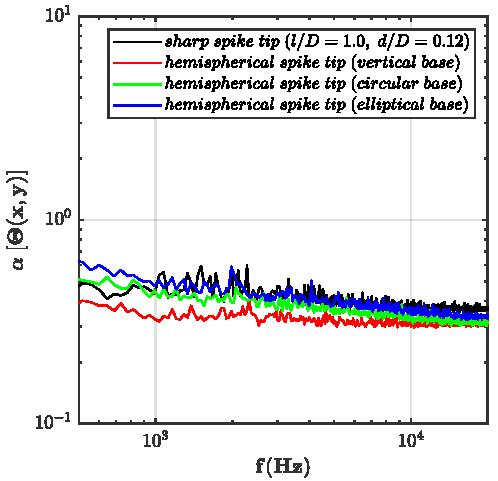}{} 
	\caption{The dynamic spectra observed from the DMD analysis ($\Theta(x,y)$) for the hemispherical forebody mounted with a typical sharp tip spike and a hemispherical tip spike of different base shapes ($l/D=1.0,\;d/D=0.12$) at $M_\infty=2.0$.}
	\label{figure26}
\end{figure}

Furthermore, both POD and DMD analyses have been carried out for the case of a hemispherical body mounted with a hemispherical tip spike of different base shapes, and the dominant energetic spatial mode and dynamic temporal mode have been computed. The dominant energetic spatial modes obtained for the hemisphere mounted with a typical sharp tip spike and with the hemispherical tip spike are shown in Figure~\ref{figure25}. The elimination of the separation shock is observed, and a reduction in the intensity of the reattachment shock strength by mounting a hemispherical tip spike can be clearly seen from the dominant energetic mode shown in Figure~\ref{figure25}b-d when compared to the intensity of reattachment shock strength generated in the case of sharp tip spike (see Figure~\ref{figure25}a). The out of phase correlation between the separation shock and the reattachment shock oscillation existing in the hemispherical body mounted with a sharp tip spike is absent when changing the spike tip to a hemispherical tip of different base shapes. Furthermore, it can be observed that the dominant spatial mode is mainly concentrated around {the reattachment shock} for the modified spike tip geometry of different base shapes. The findings from the DMD analysis (see Figure~\ref{figure26}) have also been found to be consistent with the spectral contents obtained from the pressure measurements (see Figure~\ref{figure24}). The reduction in the intensities of the broadband of spectra in the range of frequencies from 1000 Hz to 5000 Hz by adopting the hemispherical tip spike of different base shapes can be observed in Figure~\ref{figure26}. This reduction in the amplitude of the broadband of frequencies indicates the reduction of the shock-related unsteadiness. For circular and elliptical base shapes, the values are higher for reasons told in the previous paragraphs. The results from POD and DMD support the observations from the qualitative and quantitative measurements.

\section{Conclusions}
Some of the major conclusions from the present study are:

\begin{enumerate}
	\item Variations in the forebody static pressure, drag coefficient ($c_d$), and shock related unsteadiness are observed by mounting a drag-reducing spike of different geometrical parameters. 
	
	\item {Increase in the spike stem diameter ($d$) results in a minor decrease in the amount of $c_d$ reduction with respect to the forebody without a spike, whereas the increase in the length ($l$) of the spike shows a significant increase in the amount of drag reduction up to a certain length ($l/D$=1.5) for a fixed $d/D$=0.12.}
	
	\item The intensity of the shock-related unsteadiness is decreased with an increase in $d/D$. With an increase in $l/D$, the intensity of the shock-related unsteadiness increases marginally up to a spike length of [$l/D$]=1.5. A further increase of the spike length up to $l/D=2.0$ does not affect the intensity, which remains approximately unchanged. The existence of the shock-related unsteadiness in each case of a sharp tip spike is verified by modal analysis. The spatial and temporal modes obtained from POD and DMD analyses show similar trends as obtained from the measured unsteady pressure spectra. The sharp tip spike with the maximum spike diameter of $d/D =0.18$ shows the minimum level of shock-related unsteadiness. However, $c_d$ values experienced for the same case is comparatively higher due to stronger reattachment shock strength as the separation point moves.
	
	\item The spike tip geometry of spiked hemispherical bodies has a major effect on drag reduction and shock-related unsteadiness. All three spike base shapes (vertical, circular, and elliptical) produce a comparable reduction in drag and shock-related unsteadiness. It has been shown that by using a hemispherical tip spike of vertical base shape, a significant reduction in drag is achieved. Also, a significant reduction in the level of pressure loading and a slight reduction in pressure fluctuations intensity has been observed. The reason for the above findings is probably due to the elimination of the separation shock upon the mounting of the hemispherical tip spike and the consequences of the larger re-circulation region and smaller structures from K-H instabilities.
	
\end{enumerate}

\textbf{Acknowledgments}: The first author thanks the Technion research funding during his doctoral studies. The authors would like to thank the help of Mark Koifman, Michael Dunaevsky, Oleg Kan, Yafim Shulman, Nadav Shefer, and David Naftali in conducting the experiments, effectively.

Declaration of Interests: The authors report no conflict of interest.

\bibliography{article}
\end{document}